\documentclass[preprint,aps,prd,superscriptaddress,preprintnumbers,nofootinbib]{revtex4-2}
\usepackage{latexsym,amssymb,amsmath,amsfonts,graphicx,float,url,mathtools,braket,xcolor}
\usepackage{empheq}
\usepackage{bbm}
\usepackage{verbatim}
\usepackage{xcolor}
\usepackage[utf8]{inputenc}
\usepackage[hidelinks]{hyperref}
\hypersetup{colorlinks=true,
    linkcolor=blue,
    filecolor=blue,
    urlcolor=blue,
    citecolor=blue,
    }

\usepackage{amssymb}

\newcommand\eea{\end{eqnarray}}
\newcommand\bea{\begin{eqnarray}}
\newcommand\ea{\end{align}}
\newcommand\ba{\begin{align}}

\newcommand\nn{\nonumber}
\newcommand\ml{\mathcal}

\newcommand\ra{\rightarrow}

\graphicspath{{figures/}}

\begin{document}

\title{Nuclear and electron scattering by neutrinos and dark matter in condensed systems}

\author{James B.~Dent} 
\email{jbdent@shsu.edu}
\affiliation{Department of Physics$,$~ Sam ~Houston~ State~ University$,$~ Huntsville$,$~ TX~ 77341$,$~ USA}

\author{Barry A.~Friedman} 
\email{phy\_baf@shsu.edu}
\affiliation{Department of Physics$,$~ Sam ~Houston~ State~ University$,$~ Huntsville$,$~ TX~ 77341$,$~ USA}

\author{Jayden L.~Newstead}
\email{jnewstead@unimelb.edu.au}
\affiliation{ARC center of Excellence for Dark Matter Particle Physics$,$ \\~School of Physics$,$~ The~ University~ of~ Melbourne$,$~ Victoria~ 3010$,$~ Australia}

\author{Subir Sabharwal}
\email{subir@uri.edu}
\affiliation{Center for Computational Research, University of Rhode Island, Kingston RI 02881, USA}

\begin{abstract}
Low-threshold dark matter detectors, in particular cryogenic detectors based on dielectric materials, are among the best tools for probing sub-GeV dark matter masses. In the coming years detectors of this type will become sensitive to solar neutrino scattering. Previous work has shown that, for dark matter scattering at very low recoil energies, one must include collective excitations of the electrons in the solid. In this work, we have computed the collective excitations due to neutrino scattering on electrons and nuclei. We find the full electron-scattering response at leading order is captured by 5 structure factors and identify the leading component with the electron energy-loss function. Then, using silicon and germanium detectors as an example, we perform a dark matter sensitivity study and compute their respective neutrino floors. Lastly, we show that these detectors are sensitive to unexplored scenarios of beyond-Standard Model neutrino physics, within the exposure required to reach the neutrino floor.
\end{abstract}

\maketitle
\newpage 

\section{Introduction}

The direct search for dark matter producing a scattering signal in terrestrial lab-based experiments continues apace with increasing sensitivity to lighter dark matter masses and lower interaction cross sections between dark matter and Standard Model (SM) particles such as electrons and nucleons (for recent overviews of current and future detectors, see for example~\cite{LZ:2019sgr,Settimo:2020cbq,Rau:2020abt,DiGangion:2021thw,Edelweiss:2022bzh,Aalbers:2022dzr,Akerib:2022ort,Essig:2022dfa,Liu:2022zgu,Bottino:2022zky,SABRE:2022twu}). With lighter masses comes the need for lower energy thresholds, due to straightforward kinematic considerations\footnote{Alternatively, incident dark matter may have been boosted to higher energies (allowing for larger momentum transfer and therefore detection of lighter particles) through mechanisms such as annihilation in the dark sector~\cite{Agashe:2014yua,Berger:2014sqa,Kong:2014mia,Cherry:2015oca,Kim:2016zjx,Giudice:2017zke,Super-Kamiokande:2017dch,McKeen:2018pbb}, cosmic-ray interactions~\cite{Cui:2017ytb,Bringmann:2018cvk,Cappiello:2018hsu,Ema:2018bih,Alvey:2019zaa,Cappiello:2019qsw,Dent:2019krz,Krnjaic:2019dzc,Bondarenko:2019vrb,Wang:2019jtk,Plestid:2020kdm,Guo:2020drq,Ge:2020yuf,Guo:2020oum,Xia:2020apm,Dent:2020syp,Harnik:2020ugb,Ema:2020ulo,Jho:2021rmn,Das:2021lcr,Bell:2021xff,Feng:2021hyz,Xia:2021vbz,Wang:2021jic,Bardhan:2022bdg,Maity:2022exk,Bell:2023sdq,Dutta:2024kuj}, or solar reflection~\cite{An:2017ojc,An:2021qdl,Emken:2021lgc,Liang:2021zkg,CDEX:2023wfz,Emken:2024nox}.}. Technological improvements in detector design, and the implementation of new search channels are allowing access to lower-mass regions than ever before, providing further motivation for the study of possible new interaction signatures and detector responses in order to properly characterize dark matter searches. Along with new search channels in liquid noble detectors \cite{XENON:2019gfn,XENON:2019zpr},  studies of the condensed matter physics of solid-state detectors, necessary for a faithful translation from nuclear or electron interactions to detection signals, have gained in importance and popularity \cite{Trickle:2019nya,Griffin:2019mvc,Trickle:2020oki,Knapen:2020aky,Kurinsky:2020dpb,Kahn:2021ttr,Knapen:2021bwg,Knapen:2021run,Griffin:2021znd,Mitridate:2022tnv}. In addition to studying dark matter interactions in condensed-matter systems, new backgrounds that may arise from similar physical considerations also require scrutiny in order to properly isolate a new-physics signal. 

Excitation of electronic modes, either from direct scattering with electrons, or from secondary excitations due to nuclear interactions, is an important possible discovery avenue for dark matter interactions~\cite{Essig:2011nj,Essig:2015cda,Lee:2015qva,Emken:2017erx,Ibe:2017yqa,Dolan:2017xbu,Catena:2019gfa,Emken:2019tni,Bell:2019egg,Essig:2019xkx,Baxter:2019pnz,Bell:2021ihi}. Indeed, it has been shown by numerous experimental collaborations -- utilizing both solid-state and liquid noble detection methods -- that electron scattering and electronic recoils induced by nuclear scattering can be a sensitive probe for sub-GeV mass dark matter~\cite{Essig:2012yx,SuperCDMS:2018mne,DarkSide:2018ppu,DAMIC:2019dcn,SENSEI:2020dpa,XENON:2019gfn,XENON:2019zpr,XENON:2020rca,EDELWEISS:2020fxc,PandaX-II:2021nsg}. Modeling the transition from single-particle interactions to detector response modes, including the effects of collective excitations of the detector medium, has become an integral element of the direct detection analysis pipeline. 

In solid-state detectors, which will be the focus of this work, a scattering event will perturb the electron density in the detector medium. The response of the medium is encoded by a dynamic structure factor, whose leading contribution, for example, is related to the dielectric response function~\cite{mah00,alma991015472689704336,Knapen:2021run}. Incorporation of an accurate description of the detector response utilizing these elements is key to a robust analysis of predicted scattering rates involving electronic excitations, whether they are produced by particle interactions within or beyond the SM. The dielectric response can harbor interesting features such as plasmon resonances that can possibly amplify the scattering rate, thereby enhancing detection prospects~\cite{Kurinsky:2020dpb,Kozaczuk:2020uzb}. 

Similar condensed matter considerations must be included in calculations of backgrounds at such searches, as well. For example, it is well known that neutrino scattering can produce a background at upcoming direct detection searches, with the first indications of solar neutrino detection being recently announced~\cite{XENON:2024ijk,PandaX:2024muv}, along with some initial phenomenological implications of these results~\cite{AristizabalSierra:2024nwf,Li:2024iij,Maity:2024aji,DeRomeri:2024iaw,DeRomeri:2024hvc}. This background has gained the appellation `neutrino floor' or `neutrino fog', due to both the difficulty inherent in experimentally reducing scattering from neutrino sources (solar neutrinos will tend to be the dominant neutrino background for dark matter masses $\lesssim\mathcal{O}(10)$ GeV, while atmospheric neutrinos will be the leading neutrino background at higher masses), and their excellent mimicry of WIMP dark matter scattering, being neutral and weakly interacting~\cite{Billard:2013qya,Dent:2016iht,Dent:2016wor,McCabe:2017rln,OHare:2016pjy,Essig:2018tss,Wyenberg:2018eyv,Papoulias:2018uzy,Boehm:2018sux,OHare:2020lva,OHare:2021utq,Blanco-Mas:2024ale}. The background from low-energy neutrino-electron scattering and its associated neutrino fog has been explored in~\cite{Carew:2023qrj}, however a detailed treatment of the electronic excitations from neutrino scattering in condensed matter systems warrants further investigation\footnote{The possibility of neutrino scattering as the instigator of the widely reported low energy excess seen at a variety of semiconductor experiments has been studied recently~\cite{Abbamonte:2022rfh}, where it was shown to be ruled out as the progenitor of the excess.}.

In this work, we examine neutrino induced electronic excitations not only as a background to dark matter searches, but as a signal themselves. This includes a calculation of the aforementioned solid-state structure factor for SM neutrino-electron interactions. For experiments with non-negligible neutrino event rates, this framework would also provide an interesting probe of neutrino physics, akin to the plenitude of studies describing the utility of direct detection experiments for neutrino detection~\cite{Cerdeno:2016sfi,Bertuzzo:2017tuf,Dutta:2017nht,AristizabalSierra:2017joc,Huang:2018nxj,Newstead:2018muu,Dutta:2019oaj,Bell:2019egg,AristizabalSierra:2019ykk,DARWIN:2020bnc,Boehm:2020ltd,AristizabalSierra:2020zod,Khan:2020csx,Vahsen:2020pzb,Newstead:2020fie,Suliga:2020jfa,Gann:2021ndb,AristizabalSierra:2021kht,Schwemberger:2022fjl,Aalbers:2022dzr,DeRomeri:2024dbv}). 

We employ the Foldy-Wouthuysen transformation approach to the non-relativistic expansion of the interaction Hamiltonian~\cite{PhysRev.78.29,Bjorken:100769} and then calculate the neutrino scattering rates following the Kramers-Heisenberg approach~\cite{schulke2007}. We find that the neutrino-electron interaction induces five distinct `structure factors' at leading order, that depend only on the properties of the solid and these structure factors (in addition to the energy distribution of the neutrino flux) completely govern the dynamics of the scattering process. The novelty of this approach is that it provides a systematic derivation of the structure factors that are excited by a low-energy interaction. Some of the structure factors are spin-independent and depend only on the electron density in the solid whereas others depend on the magnetic properties of the solid. In this paper, we only consider solids where the spin-independent structure factor (which is related to the dielectric function of the solid) dominates the scattering cross section. However, this leaves open for future consideration studies of interactions in solids whose magnetic properties furnish the leading-order contribution. Such collective modes involving spin waves, known as magnons, have been considered as a detection mode for light spin-dependent dark matter ~\cite{Trickle:2019ovy,Trickle:2020oki,Esposito:2022bnu,Marocco:2025eqw}.

In this work, we then make use of the dielectric response function as laid out in~\cite{Kozaczuk:2020uzb,Knapen:2020aky,Knapen:2021run}\footnote{See also~\cite{Dreyer:2023ovn} for an ab-initio calculation for dark matter-electron scattering in crystals.} in order to describe both direct neutrino-electron interactions from solar neutrinos, with $pp$ chain neutrinos being the dominant component, and electron signals induced by neutrino-nuclear interactions in a Migdal-effect style process.

We find that electronic excitations from incident solar neutrinos occur at the rate of $\sim 10^{-5}$/kg/yr below 1 keV in germanium and silicon based detectors. This is an irreducible background, covering the whole kinematic range of DM-electron scattering. This allows us to calculate a `neutrino-floor' region where such neutrino-induced backgrounds limit the sensitivity that experiments can achieve. Additionally, we detail the prospects of detecting neutrinos with such devices, including a generic non-standard interaction (NSI) model of neutrino-SM interactions through a light scalar mediator. We find that detectable neutrino scattering can occur in an interesting region of the dark matter cross section vs. mass parameter space that can be reached with near-term detector technology.

The remainder of the paper is as follows: Section~\ref{sec:nuinducedelectronicexcitations} provides the details of our scattering rate calculation for electronic excitations from Standard Model $\nu$-$e^{-}$ interactions and nuclear recoil induced production. The resulting observable rates of ionization in silicon and germanium targets is presented in Sec.~\ref{sec:results}. Results of the sensitivity studies are presented and discussed in Sec.~\ref{sec:sensitivity}, with concluding remarks in Sec.~\ref{sec:summary}. We provide our conventions in App.~\ref{app:conventions}, followed by detailed calculations of the structure factors and rates in App.~\ref{app:enuinteractions} and App.~\ref{app:brem}. App.~\ref{app:enuinteractions} also includes details on novel solid responses which are distinct from the canonical density-density correlators (represented in the dielectric function).

\section{Neutrino induced electronic excitations}
\label{sec:nuinducedelectronicexcitations}

In this section we describe the rates for collective electronic excitations via direct Standard Model neutrino-electron interactions, and through nuclear recoils (similar to the Migdal effect). Lastly, we discuss beyond Standard Model neutrino-electron interactions mediated by a hypothetical light particle (non-standard neutrino interactions). We provide the transition probability of the scattering process in terms of the dynamic structure factors, of which we have found five at leading order in $1/m_{e^{-}}$, where $m_{e^{-}}$ is the mass of the electron. These structure factors encapsulate crystal properties, and are somewhat analogous to the different nuclear responses familiar from the effective field theory treatment of dark matter direct detection~\cite{Fitzpatrick:2012ix}.

\subsection{Direct production via neutrino-electron scattering}

We start with the low-energy effective interaction Lagrangian between a charged lepton and a neutrino, which is given by the V-A Fermi term,
\bea
\ml{L}_{int} = -\frac{4G_F}{\sqrt{2}}\left[\bar{\nu}\gamma_\mu\left(\frac{1-\gamma^5}{2}\right)\nu\right]\left[\bar{\psi}\gamma^\mu\left(g_L\frac{1-\gamma^5}{2}+g_R\frac{1+\gamma^5}{2}\right)\psi\right]+h.c.,
\label{eq:lagrangian}
\eea
where $\psi$ and $\nu$ are the charged lepton (which will be the electron for our purposes) and  neutrino field respectively, $h.c.$ denotes the hermitian conjugate, and the couplings are real and depend on whether the interaction is mediated by a neutral boson ($Z^0$) or charged vector boson ($W^\pm$) exchange. Summing over both exchange modes leads to the couplings to electons  
\bea
g_L &=& \left(-\frac{1}{2}+\sin^2\theta_w\right)_{Z^0} + \left(1\right)_{W^\pm}=\frac{1}{2}+\sin^2\theta_w\nn\\
g_R &=& \left(\sin^2\theta_w\right)_{Z^0} + \left(0\right)_{W^\pm}=\sin^2\theta_w,
\eea
where $\theta_w$ is the Weinberg angle and the $W^\pm$ contribution is included for interactions between $\nu_e$ and electrons, and is absent for $\nu_{\mu}$ and $\nu_{\tau}$ interactions with electrons. Expanding the interaction Hamiltonian operator systematically in powers of the inverse electron mass to obtain higher order non-relativistic terms 
(details can be found in Appendix~\ref{app:foldy}) we get
\bea
\hat{\ml{H}}^{(3)}_{e^{-}} &\approx& \beta\left[m_e-\frac{\left[\nabla^2+2i(\vec{\ml{V}}\cdot\vec{\nabla})+i(\vec{\nabla}\cdot\vec{\ml{V}})+\vec{\sigma}\cdot\left(\vec{\nabla}\times\vec{\ml{V}}\right)\right]+i \vec{\sigma}\cdot\left[2\ml{A}\ \vec{\nabla}+ (\vec{\nabla}\ml{A})\right]}{2m_e}\right]\nn\\
&&-\beta\ \frac{p^4}{8m_e^3}+\left[\left(\frac{g_R+g_L}{g_R-g_L}\right)\ml{A}-\left(\frac{g_R-g_L}{g_R+g_L}\right)(\vec{\sigma}\cdot\vec{\ml{V}})\right]\\
&&-\frac{1}{8m_e^2}\left[\left(\frac{g_R+g_L}{g_R-g_L}\right)\nabla^2 \ml{A}-\left(\frac{g_R-g_L}{g_R+g_L}\right) \nabla^2(\vec{\sigma}\cdot\vec{\ml{V}})\right]\nn\\
&&-\frac{i}{8m_e^2}\left[-\vec{p}\cdot\partial_t\vec{\mathcal{V}} +i\vec{\sigma}\cdot\left(2\partial_t\vec{\mathcal{V}}\times\vec{p}-\vec{p}\times\partial_t\vec{\mathcal{V}}-i\vec{p}\,\dot{\ml{A}}\right)\right]\nn,
\eea
where $\vec{p}$ is the usual momentum operator and we use the shorthand $\ml{A} \equiv ({G_F}/{\sqrt{2}}) \ (g_R-g_L)\ \left(V^0-A^0\right)$ and $\ml{V} \equiv ({G_F}/{\sqrt{2}})\ \left(g_R+g_L\right) \left(\vec{V}-\vec{A}\right)$, where $V_\mu=(V^0, \vec{V})$ and $A_\mu=(A^0, \vec{A})$ are the neutrino vector and axial covariant currents respectively. The transition probability for the process $e^{-}+\nu_{e^{-}}\ra e^{-}+\nu_{e^{-}}$, where $e^{-}$ is an electron in the solid, impinged upon by an incident neutrino can be calculated using Fermi's golden rule,
\bea
\Gamma = {2\pi}\ \Big|\bra{\vec{p}_{\nu, i};i}\hat{H}_{e^-}\ket{\vec{p}_{\nu, f};f}\Big|^2\delta(E_{e,i}-E_{e,f}+E_{\nu, i}-E_{\nu,f}),
\eea
where $(E_{\nu,i},\ \vec{p}_{\nu,i})$ and $(E_{\nu,f},\ \vec{p}_{\nu,f})$ are the incident and scattered energy and momentum of the neutrinos, and $i$ and $f$ are the initial and final electronic states of the electrons in the solid, respectively. Using only the leading order `even term' (see Appendix~\ref{app:transitionprobability}), we arrive at the following expression for $\Gamma$,
\bea
\label{eq:M_NR1}
\Gamma &=&\left(\frac{4G_F^2}{E_{\nu,i}E_{\nu,f}V^2}\right)\left[g_{+}^2\left(E_{\nu,i}E_{\nu,f}+\vec{p}_{\nu,i}\cdot\vec{p}_{\nu,f}\right) S_1+ 2 g_{+}g_{-}\left\{E_{\nu,f}\ (\vec{p}_{\nu,i}\cdot\vec{S}^\sigma_2)+E_{\nu,i}\ (\vec{p}_{\nu,f}\cdot\vec{S}^\sigma_2)\right\}\right.\nn\\
&+&\left.g_{-}^2\left(E_{\nu,i}\ (\vec{p}_{\nu,f}\cdot\vec{S}^\sigma_4)-E_{\nu,f}\ (\vec{p}_{\nu,i}\cdot\vec{S}^\sigma_4)+2\ p_{\nu,i}^k\ p_{\nu,f}^l\ S^{kl,\sigma}_5+ (E_{\nu,i}E_{\nu,f}-\vec{p}_{\nu,i}\cdot\vec{p}_{\nu,f})\ S^\sigma_3\right)\right]
\eea
where the `structure factors' $\{S\}$ depend solely on the electronic properties of the solid and are typically calculated numerically, and $g_{\pm}=g_R\pm g_L$. Here $V$ is the fiducial quantization volume which cancels out when calculating any physical quantity. The differential rate, $\ml{R}$ as a function of the energy deposited in the solid, $\omega\equiv E_{\nu,i}-E_{\nu,f}$, can be calculated from the transition probability as (see Appendix~\ref{app:transitionprobability} for derivation),
\bea
\label{eq:direct_rate}
\frac{d\ml{R} }{d\omega}=\int_{E_{\nu,\rm{min}}} \frac{dE_\nu,i}{E_\nu,i} \frac{d\Phi_\nu}{dE_\nu,i}\left[\int_{k_{min}}^{k_{max}} \frac{dk}{2\pi}\ k \Gamma' \right](E_{\nu,i}-\omega)
\eea
where $k_{min}$ and $k_{max}$ are the kinematic boundaries of the momentum transferred, $\vec{k}$ to the solid, $\vec{k} = \vec{p}_{\nu, i} - \vec{p}_{\nu, f}$. Here $\Gamma'=V^2 \Gamma$ (thus producing the cancellation of $V$) and $\Phi_\nu(E_{\nu,i})$ is the flux of the incident neutrinos.\\

Since the targets of our study are silicon and germanium, which are not spin-polarized, we will neglect the contributions of $\{S_{2,..,5}\}$ which depend on the electron spin density (however, there may be induced interactions where the leading contribution comes from these magnetic terms. We leave such explorations to future work). As a result, the $S_1$ factor is the most relevant structure factor for our analysis. 

Further, it can be analytically shown that the $S_1$ structure factor is equal to the Fourier transform of the electronic density correlations. This, in turn, is directly related to the dielectric response of the solid through the relation
\bea
\label{struct_dielectric}
\frac{S_1(\omega, \vec{k})}{V}=\left(\frac{k^2}{4\pi^2e^2}\right)\Im\left[-\frac{1}{\epsilon(\omega, \vec{k})}\right]
\eea
where $\epsilon(\omega, \vec{k})$ is the total dielectric constant and $\Im[...]$ is the imaginary part of its argument. Note that our $S_1$ is defined in a consistent fashion with the normalization convention in~\cite{schulke2007} where the dynamic structure factor scales with the volume $V$, and thus is an extensive system property. This is different from the convention in~\cite{Knapen:2021bwg} for example, where the dynamic structure factor is defined with $1/V$ scaling so that it is already an intensive property of the system.

It's interesting to note some differences between Eq.~\eqref{eq:M_NR1} and the expression for the traditional electron-electron case  as seen, for example, in classic works such as~\cite{PhysRev.57.485,BrownRavenhall,allison1980a}, and measured  in electron energy loss spectroscopy (EELS)~\cite{schulke2007}. In the EELS case, the interaction which is mediated by photons can be analyzed in the Coulomb gauge where one finds that two terms contribute at the tree level -- one is a density-density (charge) term owing to the Coulomb potential whereas the other is the current-current interaction due to photon exchange. However, since both of these are $\ml{O}(\alpha)$, they contribute at the leading order.\footnote{This is also the case in some EELS-type extensions like those in some recent works~\cite{Essig:2024ebk}.} On the other hand, the neutrino case presents a contrast because $S_1$ and the four other electron spin-dependent responses $\{S_2^\sigma...S_5^\sigma\}$ are all induced at leading order in the effective coupling constant, $G_F^2$. As a result, the strength of contribution of these spin-dependent responses is dictated by material properties and not by the order in perturbation theory (because these four spin-dependent terms are already present at the leading order).

\subsubsection{Scattering Kinematics}

We will want to compare the rate induced by neutrino scattering to that of dark matter-electron scattering. As we are considering non-boosted and, therefore, non-relativistic galactic dark matter, along with relativistic neutrinos, it is of interest to examine the different kinematic regimes that arise in their respective scattering processes. The disparate nature of these scattering processes in terms of energy and momentum transfer will be shown to probe different regions of the response function, including differential access to the plasmon resonance area.

For non-relativistic dark matter of mass $m_\chi$ and incident velocity, $\vec{v}_{\chi,i}$, scattering off an electron in its rest frame, the energy, $\omega$, gained by the electron in the scattering process is
\bea
\omega &=& \left(E_{\chi,i}-E_{\chi,f}\right)
\eea
where $E_{\chi,i}$ and $E_{\chi,f}$ are the initial and final dark matter energies, respectively. Here we have assumed that any atomic recoil is negligible. Using the non-relativistic relations for the energy terms, we find
\bea
\omega = \frac{m_\chi v^2}{2}-\frac{|m_\chi\vec{v} - \vec{k}|^2}{2m_\chi} ,
\eea
where $\vec{k}$ is the momentum transferred to the electron in the interaction. Thus, we can find the minimum incident velocity needed to produce an electronic energy of $\omega$ at a momentum transfer of magnitude $k \equiv|\vec{k}|$
\bea
v_{min} = \frac{\omega}{k}+\frac{k}{2m_\chi}
\eea
Additionally, we can solve for the kinematic boundaries of the momentum transfer, where we find
\bea
k = m_\chi |\vec{v}|\cos\theta \pm \sqrt{(m_\chi|\vec{v}|\cos\theta)^2 - 2m_\chi\omega},
\eea
where $\theta$ is the angle between the incident dark matter velocity, $\vec{v}$, and the momentum transfer $\vec{k}$. We can see that the maximum momentum transfer occurs at
\bea
k_{max} = m_{\chi}v_{max} + \sqrt{(m_{\chi}v_{max})^2 - 2\omega m_{\chi}},
\eea
where the maximum speed is provided when the dark matter escape velocity, $\vec{v}_{esc}$, and the earth's velocity, $\vec{v}_{E}$, are aligned, $v_{\max} = |\vec{v}_{esc}| + |\vec{v}_{E}|$. 

For neutrino-electron scattering, the relativistic nature of the neutrino provides a different situation. The four-momentum of the neutrino is $p^{\mu}_{\nu,i} = (E_{\nu,i}, \vec{p}_{\nu,i})$, where $|\vec{p}_{\nu,i}| \simeq E_{\nu,i}$ holds in the relativistic limit. The electronic energy is given by the difference in initial and final neutrino energies
\bea
\omega &=& E_{\nu,i} - E_{\nu,f}
\\
&=& E_{\nu,i} - \sqrt{E_{\nu,i}^2 + k^2 - 2E_{\nu,i}k\cos\theta},
\eea
where, as in the dark matter case, $k\equiv|\vec{k}|$, is the magnitude of the transferred momentum, and $\theta$ is the angle between the incident and final neutrino three-momenta. From this relation we can find the minimum neutrino energy required to provide an energy of $\omega$ to the electron with momentum transfer of $k$
\bea
E_{\nu,min} = \frac{k+\omega}{2} 
\eea
Alternatively, one can describe the relation between $k$ and $\omega$ as
\bea
k = E_{\nu,i}\cos\theta \pm \sqrt{(E_{\nu,i}\cos\theta)^2 - 2E_{\nu,i}\omega + \omega^2}
\eea
which provides the kinematic boundaries of interest for neutrino-electron scattering.

\begin{figure*}[t]
    \centering
    \includegraphics[width=8cm]{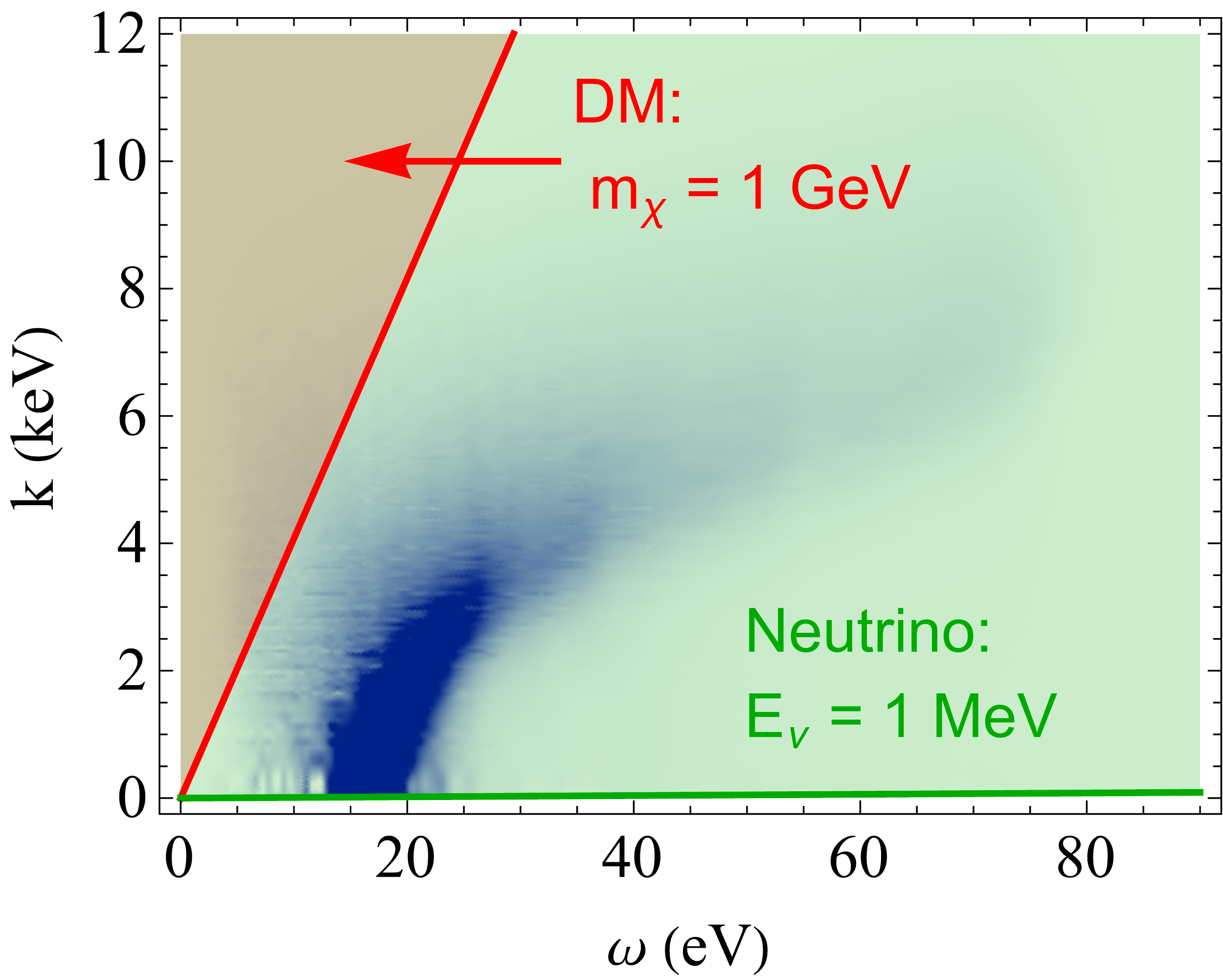}
    \caption{Region of phase space integration for non-relativistic DM (the region above and to the left of the red line) and neutrino (green) compared to electron energy-loss function $\Im(-{1/\epsilon})$ of silicon (shaded blue). The visible part of this function corresponds to the plasmon resonance.}
\label{fig:dielectricF}
\end{figure*}

\subsection{Nuclear recoil induced production}

In addition to direct neutrino-electron scattering, one can also consider electronic energy deposition from an associated neutrino-nucleus scattering process. Analogous electron ionization mechanisms have been examined for dark matter-nucleus scattering which has included plasmon production~\cite{Kurinsky:2020dpb,Kozaczuk:2020uzb}, and the Migdal effect~\cite{Ibe:2017yqa,Dolan:2017xbu,Bell:2019egg,Baxter:2019pnz,Essig:2019xkx,Liu:2020pat,Knapen:2020aky,Liang:2020ryg,Bell:2021zkr,Acevedo:2021kly,Liang:2019nnx,GrillidiCortona:2020owp,Herrera:2023xun}
 (\cite{Bell:2019egg} also examined the Migdal effect in the context of neutrino-nucleus scattering for liquid noble detectors). The rates for these secondary production mechanisms are typically suppressed by orders of magnitude relative to the primary coherent and elastic nuclear scattering process (see also the reduction for photon bremsstrahlung in dark matter scenarios~\cite{Kouvaris:2016afs}).

In this subsection we will briefly describe electronic excitations from neutrino-nucleus scattering which includes the effects of the response of the detector material via the energy loss function. The scattering process is effectively a 2-to-3 interaction that will factorize into the coherent and elastic 2-to-2 differential cross-section in the reference frame of a nucleus initially at rest~\cite{Freedman:1973yd}
\bea
\label{eq:CEvNS}
\frac{d\sigma_{\nu N}}{dE_R} = \frac{G_F^2m_T}{4\pi}Q_V^2\left(1-\frac{m_TE_R}{2E_\nu^2}\right)F^2(k^2),
\eea
where $E_R$ is the nuclear recoil energy, $G_F$ is Fermi's constant, $m_T$ is the mass of the target nucleus, $F(k^2)$ is the nuclear form factor, $Q_V = N-(1-4\sin^2\theta_W)Z$ is the vector charge of the nucleus that depends on the number of neutrons, $N$, the weak mixing angle, $\theta_W$, and the number of protons, $Z$, along with terms that account for the additional electronic production mode. These terms have the effect of kinematically suppressing the overall process relative to the 2-to-2 scenario by several orders of magnitude. The appearance of such a suppression is common to both the bremsstrahlung and Migdal processes.

Neglecting the sub-leading structure factors, we can derive  the double differential scattering cross-section for plasmon production (see Appendix \ref{app:brem} for a derivation),
\bea
\label{eq:nuclearplasmon}
\frac{d^2\sigma}{dE_Rd\omega}=\frac{4\alpha}{3\pi^2}\frac{E_R}{m_T}\int dk \frac{k^2 Z_{\rm{ion}}^2(k)}{\omega^4}\left[\frac{d\sigma_{\nu N}}{dE_R}\right]\Im\left[-\frac{1}{\epsilon(\omega, \vec{k})}\right]
\eea
where $Z_{\rm{ion}}^2(k)$ is the Fourier transformed potential due to the charge of the nucleus and core-electrons, and as before, $\epsilon(\omega, \vec{k})$ is the Fourier-space dielectric constant.
  
\subsection{Light Mediators in Neutrino Interactions}

Along with the Standard Model neutrino-electron and neutrino-nucleus processes, one can introduce new neutrino interactions mediated by particles beyond the Standard Model. The study of novel neutrino interactions has become a vast and quickly evolving field, and we point the interested reader to the papers~\cite{Farzan:2017xzy,Proceedings:2019qno} for an overview of the topic. These interactions are known as `non-standard interactions' (or NSI) when in the context of effective field theories of mediators significantly more massive than the typical momentum transfer of the scattering process of interest. For our purposes, we will work with new light mediators whose mass is small compared to the $\sim$MeV momentum transfers typical of solar neutrinos. This will allow us to explore the possible alterations to the electronic production rate in semiconductors in a generic light-mediator framework (in~\cite{Chao:2021ahl}, plasmon production from neutrino scattering within such a scenario was examined in the context of exploring the cosmic neutrino background). The inclusion of light mediators tends to enhance the scattering rate at low energies, which is precisely the regime in which we are interested.

For the case of neutrino-electron scattering, we adopt a vector mediator model with interactions provided by the interaction Lagrangian $\mathcal{L}_{\phi,int} = Z_{\mu}'(g_{\nu,Z'}\bar{\nu}_L\gamma^{\mu}\nu_L + g_{ \ell,Z'}\bar{\ell}\ell)$, where the vector has a mass $m_{Z'}$ and $\ell$ represents the charged lepton states\footnote{See, for example,~\cite{Cerdeno:2016sfi,Dent:2016wcr,Dent:2019krz} for details of the cross-section in this type of model.}. 
For neutrino-nucleus scattering we will introduce not only a vector mediator, but also a scalar and axial-vector mediator with couplings at the nucleon level. These will augment the SM scattering described in Eq.(\ref{eq:CEvNS}). For details regarding the differential cross-sections for the electron and nuclear scattering with the inclusion of these NSI, see~\cite{Cerdeno:2016sfi,Dent:2016wcr}.

Having elucidated the various scattering processes of interest, we now turn to an examination of their rates and detection prospects in the context of current and future experiments.

\section{Experimental scattering rates}
\label{sec:results}

In this section we will summarize the calculation of scattering rates for the various processes that are of interest for low-threshold DM and neutrino detectors that operate through the detection of ionization. We will then proceed to calculate the potential these detectors have to discover (or constrain) DM and new light mediators in the neutrino sector.

Some DM detectors have the ability to discriminate nuclear and electronic recoils. In liquid-noble time-projection chambers this is achieved through the ratio of prompt and delayed scintillations, or pulse-shape analysis. For cryogenic detectors like the interleaved z-sensitive ionization and
phonon detectors (iZIPs) used by SuperCDMS, discrimination is performed with the ratio of the charge and phonon signals of a given event. 

Other detector types, such as the CDMSlite high voltage (HV) detectors exchange the ability to discriminate between electron and nuclear recoils for a lower threshold reach. In these detectors, a bias voltage is applied which amplifies the ionization signal through the Neganov-Trofimov-Luke effect~\cite{Neganov:1985khw,Luke:1988xcw}, swamping the small nuclear recoil phonon signal. Recently, hybrid HV detectors have been produced which aim to combine a low threshold reach through HV amplification with the ability to discriminate between electron and nuclear recoils. These detectors have a bulk fiducial region with low bias voltage and a `neck' with HV, that can potentially allow discrimination down to 500 eV NR~\cite{Neog:2020ily,Maludze:2024rub}.

In this work, we will not assume a particular technology beyond sensitivity to ionization. Instead we make the idealized assumption that a future detector will be able to count electrons with perfect accuracy and be sensitive to a single electron.

\subsection{The Migdal effect}

The nuclear scattering processes, CEvNS and DM-nucleus, can directly induce ionization via the Migdal effect. This occurs because, assuming the impulse approximation in the frame of the nucleus, the electron cloud receives an instantaneous boost. The wavefunction of the boosted electrons now have a non-zero overlap with unbound (and higher bound) states and  thus there is a small probability of ionization (or excitation) occurring.  These probabilities are typically calculated assuming isolated atoms~\cite{Ibe:2017yqa,Cox:2022ekg}.

In a condensed matter context, the atom itself is sitting in a potential and the impulse approximation can be applied so long as the time scale of the collision ($\sim 1/E_N$) is much faster than the time taken for the atom to traverse the lattice potential ($1/\bar\omega$)~\cite{Knapen:2020aky}. In the case of DM scattering, this is valid for $m_\chi\gtrsim70$ MeV. Additionally, the isolated atom approximation is no longer suitable due to the material's band structure. This is most pronounced for the electrons occupying the valence band which can be more easily liberated into the conduction band. The orbitals of the core electrons are less affected, and while there is a small modification due to ionization into the conduction band, we neglect that effect here and treat them with the isolated atom approximation.

\subsection{The dielectric function}

For our rate calculations involving valence electrons (e.g. Eqs.~(\ref{eq:direct_rate}) and~(\ref{eq:nuclearplasmon}) above) we need to know the dielectric function, $\epsilon(\omega,\vec{k})$, for the target materials. This can be calculated with a model, e.g. the Mermin method~\cite{PhysRevB.1.2362,Knapen:2021bwg}, or with density-functional theory (DFT). The Mermin method is a phenomenologically motivated extension to the Lindhard model~\cite{PhysRevB.1.2362}. It is fitted with data from the $k = 0$ (optical) limit and extrapolated from there. It also fails at very low $\omega$ because it doesn't model the detailed band structure~\cite{Knapen:2021bwg} very well. Conversely, DFT is an ab initio method for calculating the band structure of materials from the periodic potential of the crystal. In this work we choose the DFT method because it's more accurate in our energy region of interest~\cite{Knapen:2021run}.\\

Specifically, in this work we will make use of the dielectric functions calculated in \cite{Knapen:2021bwg,Knapen:2021run}, where the DFT code GPAW was used~\cite{PhysRevB.71.035109,45bf832b43e2475cae28dd9980410f42}. These results have been averaged over direction and thus only depend on the magnitude of $\vec{k}$. The results of this calculation are shown in Fig.~\ref{fig:dielectricF} along with example regions of phase space integration for DM and neutrino scattering. The essentially massless neutrinos, being relativistic, have very different kinematics from heavy, non-relativistic DM. For a given $\omega$, the neutrino phase space extends to much lower $k$ than does that of DM, even encompassing the plasmon resonance.

\subsection{Neutrino scattering rates}

\begin{figure*}[t]
    \centering
    \includegraphics[width=0.45\textwidth]{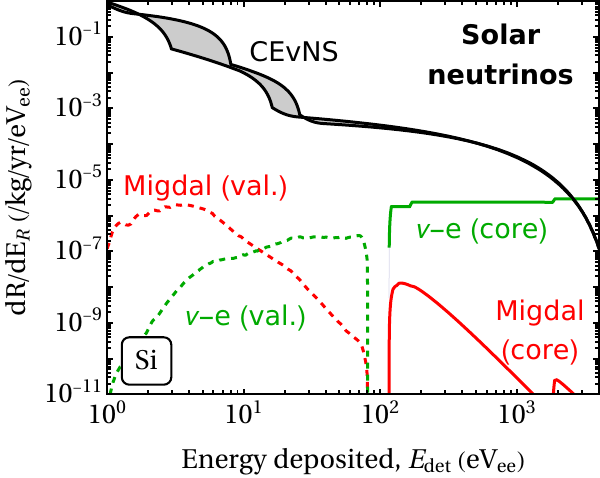}\hspace{5mm} 
    \includegraphics[width=0.45\textwidth]{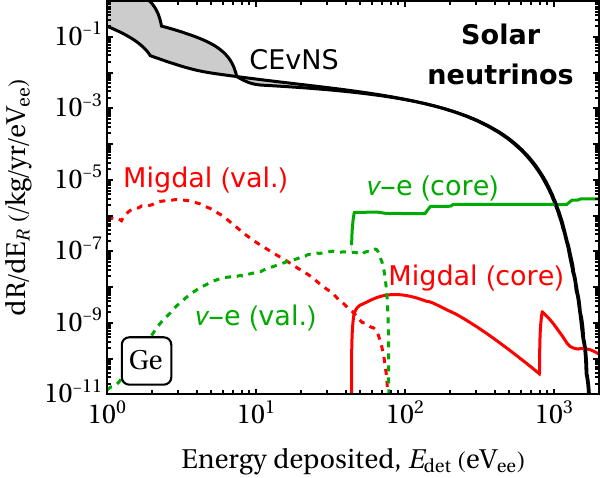}\\
    \caption{Summary of neutrino-electron and neutrino-nucleus scattering rates in silicon (left) and germanium (right). The uncertainty bands on the nuclear recoils are derived from the different quenching models.}
    \label{fig:nuRates}
\end{figure*}

For sufficient exposures, DM direct detection experiments eventually encounter an irreducible neutrino background, either from neutrino-electron scattering or coherent elastic neutrino-nucleus scattering (CEvNS). For experiments with thresholds below a few keV, the most prominent background is the solar neutrino background induced by CEvNS. At higher energies CEvNS from atmospheric neutrinos, and electron scattering by solar neutrinos become relevant. For low-threshold condensed matter experiments, CEvNS from solar neutrinos will start to be encountered in exposures as little as $\sim10$ kg-years. 

The differential scattering rate, per unit detector mass, for CEvNS is calculated using the differential cross section of Eq.~\eqref{eq:CEvNS}, weighted by the differential solar neutrino flux:
\begin{equation}
    \frac{d\ml{R}_{\nu N}}{dE_R} = \frac{1}{m_T} \int_{E_{\nu}^{\rm{min}}} dE_\nu \frac{d\sigma}{dE_R} \frac{d\Phi_\nu}{dE_\nu},
\end{equation}
where $m_T$ is the mass of the nuclear target, and $E_{\nu}^{\rm{min}}$ is the minimum neutrino energy required to cause a recoil of energy $E_R$. 

The scattering rate of neutrinos from electrons is much smaller than the rate due to CEvNS. However, the CEvNS recoils are concentrated in a narrower energy range, below roughly a keV. We break the neutrino-electron scattering rate calculation into two parts; at very low recoil energies (below around 100 eV) the band structure of the material becomes important. In this region, the calculation of Sec.~\ref{sec:nuinducedelectronicexcitations} will be used to describe scattering for the valence electrons. For the core electrons, we use the step approximation:
\begin{equation}
    \frac{d\ml{R}^{\rm{core}}_{\nu e}}{dE_e} = \frac{1}{m_T}\sum_{j} \int dE_\nu \Theta(E_e - E_{j})\left( P_{s}\frac{d\sigma_{\nu_e}}{dE_e} + (1-P_{s})\frac{d\sigma_{\nu_\mu}}{dE_e} \right) \frac{d\Phi_\nu}{dE_\nu}
\end{equation}
where the sum runs over all core electrons and $\Theta(E_e - E_{j})$ is a Heaviside function which enforces the step approximation, i.e. ensuring that electrons cannot be liberated with energy below their binding energy, $E_{j}$. Neutrino oscillations are taken into account with $P_{s}$, the survival probability of the solar electron neutrinos. In general, $P_{s}$ is a function of energy, but for simplicity we neglect it here and use an average value for each solar neutrino component.

We calculate the rate of ionization of valence electrons due to the Migdal effect from CEvNS with the cross section derived in Eq.~(\ref{eq:nuclearplasmon}). Neglecting the contribution to the deposited energy from the nuclear recoil, we have
\begin{equation}
    \frac{d\ml{R}^{\rm{val}}_{\nu N}}{d\omega} = \frac{1}{m_T}\int dE_\nu \frac{d\Phi_\nu}{dE_\nu} \int dE_R \frac{d^2\sigma}{dE_R d\omega}. 
\end{equation}
For the core electrons we follow the formalism of Ibe et al.~\cite{Ibe:2017yqa}, however, with ionization probabilities taken from the more detailed atomic calculations of Cox et al. \cite{Cox:2022ekg}.  

To compute the scattering rates, in this work we use the solar neutrino fluxes from~\cite{Haxton:2012wfz}. The resulting scattering rates are shown in Fig.~\ref{fig:nuRates}. Due to the energies of the solar neutrinos, the nuclear recoils deposit sufficient energy to be observable. Therefore, in detectors without discrimination, the neutrino-electron and Migdal CEvNS events would not be observable.

\subsection{DM scattering rates}

\begin{figure*}[t]
    \centering
    \includegraphics[width=0.45\textwidth]{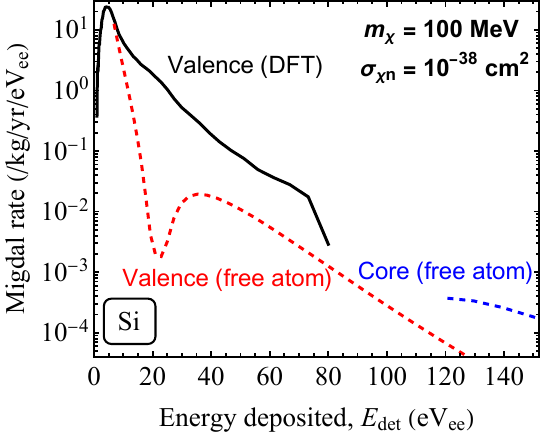}~~~~
    \includegraphics[width=0.45\textwidth]{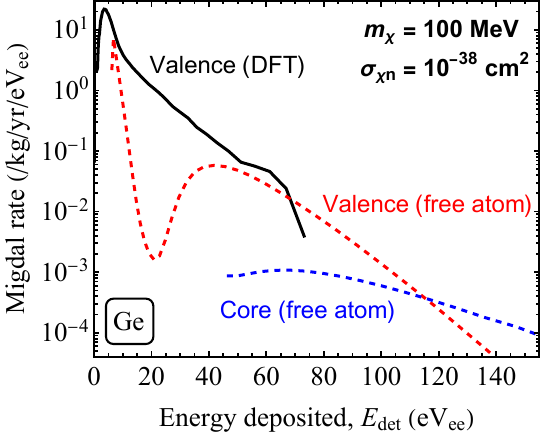}\\
    \caption{Comparison of the Migdal effect rate for valence electrons via the dielectric function (from DFT calculations) versus the free atom approximation }
    \label{fig:migComp}
\end{figure*}

Dark matter-nucleus scattering is assumed to proceed via a heavy mediator, which in the non-relativistic regime results in a contact operator. In this work we further restrict our consideration to a spin-independent interaction. The differential DM-nucleus cross section with respect to the recoil energy is
\begin{equation}
    \frac{d\sigma_{\chi  T}}{dE_R} = \frac{m_T}{2\mu_{\chi n}^2 v^2} \sigma_{n\chi} A^2 F^2(k),
\end{equation}
where $m_T$ and $A$ are the nuclear target mass and atomic number, respectively, $\mu_{n\chi}$ and $\sigma_{n\chi}$ are the DM-nucleon reduced mass and cross section, respectively, and $F(k)$ is the nuclear form factor.
The scattering rate is found by integrating over the incoming DM's velocity distribution, $f(v)$, in the lab frame:
\begin{equation}
    \frac{d\ml{R}_{\chi N}}{dE_R} = \frac{1}{m_T} \frac{\rho_\chi}{m_\chi} \int_{v_{\rm{min}}} dv \frac{d\sigma}{dE_R} f(v),
\end{equation}
where we make the assumptions of a truncated Maxwell-Boltzmann distribution with velocity dispersion $v_0 = 220$ km/s, escape velocity $500$ km/s, and local dark matter density $\rho_\chi=0.4$ GeV/cm$^3$. The velocity distribution needs to be boosted into the lab frame by $v_e = |\vec{v}_0 + \vec{v}_{\odot,\textrm{pec}}|=236$ km/s, where $\vec{v}_{\odot,\textrm{pec}}$ is the Sun's peculiar velocity and we have neglected the Earth's motion around the Sun which induces an annually modulating and directional signal.

Following~\cite{Knapen:2021run}, for dark matter-electron scattering from valence electrons (at zero temperature) we have:
\begin{equation}
    \frac{d\ml{R}^{\rm{val}}_{\chi e}}{d\omega} = \frac{1}{\rho_T} \frac{\rho_\chi}{m_\chi}\frac{\sigma_{e\chi}}{\mu^2_{\chi e}} \frac{\pi}{\alpha_{EM}}\int d^3v f_\chi(v) \int \frac{d^3k}{(2\pi)^3} k^2 F_{DM}^2(k)\  \Im\left[\frac{-1}{\epsilon_L(\omega, \vec{k})}\right]\delta\left(\omega+\frac{k^2}{2m_\chi}-\vec{k}\cdot\vec{v}\right)
\end{equation} 
where the integral also runs over the transferred momentum, $k$. We neglect the scattering of dark matter from core electrons, which will be greatly suppressed for the DM masses we consider due to the binding energy being similar to, or greater than the DM kinetic energy.

For the DM Migdal calculation we make an equivalent split to the neutrino case, with the valence calculation following Knapen et al.~\cite{Knapen:2020aky}, while the core calculation follows Ibe et al. with the Cox et al. probabilities. In Fig.~\ref{fig:migComp} we show an example Migdal rate for DM with a mass of 100~MeV. For DM of this light mass, the maximum nuclear recoil energy is below $1 \rm{eV}_{\rm{ee}}$ in both targets, which is why it does not appear on the plot. For comparison purposes only we include the free-atom calculation of valence electrons. The discrepancy between the free-atom calculation and the DFT calculation highlights the effect of the band structure on the Migdal rates at very low energies.

\subsection{Ionization rates}

\begin{figure*}[t]
    \centering
    \includegraphics[width=0.45\textwidth]{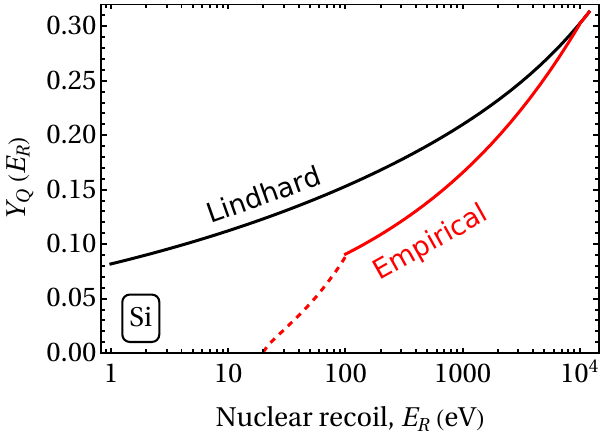}\hspace{5mm} 
    \includegraphics[width=0.45\textwidth]{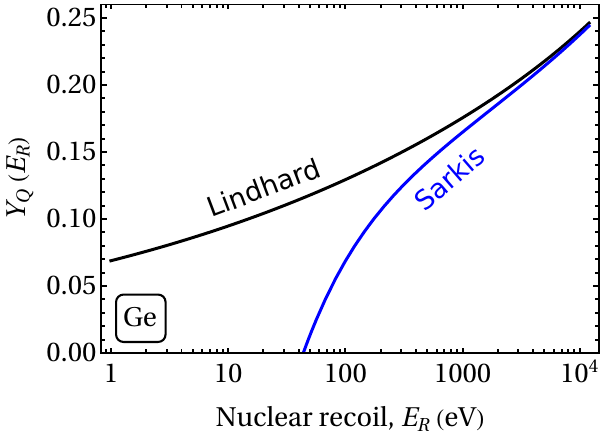}\\
    \caption{Models for nuclear recoil quenching in silicon (left) and germanium (right). In this work we use the Lindhard model as the upper bound and either the Sarkis model (germanium) or an empirical fit (silicon, from \cite{SuperCDMS:2023geu}) as the lower bound.}
    \label{fig:quenching}
\end{figure*}

To compare rates computed with the above formulae we must express the electron and nuclear recoil energies on a common scale, as is seen by a detector. Since nuclear recoil energies are quenched, i.e. significant energy is lost to heat (phonons), it's practical to use the electron recoil equivalent as the detected energy scale (e.g.,~$\rm{eV_{ee}}$). Energies are then calculated as $E_{ee} = Y_Q (E_R) E_R$. For electron scattering there is no quenching, implying $Y_Q(E_e)=1$, while for nuclear scattering the form of this function must be ascertained from a condensed-matter model, or taken from a calibration measurement. 

The Lindhard model has been a standard choice for quenching factors and provides reasonable agreement with experiment at higher energies. However, there is considerable disagreement at lower energy recoils where it underestimates the resulting quenching. Other models, such as the Sarkis model, attempt to remedy this. We will follow~\cite{Schwemberger:2022fjl} and bracket this variation by performing our calculations with two models and present the results as a band. Since~\cite{Schwemberger:2022fjl} was published, more experimental data has become available for the quenching factor of silicon. This seems to show that previous estimations of the Sarkis model~\cite{Sarkis:2020soy} were too pessimistic, therefore we will use the empirical model for silicon provided in~\cite{SuperCDMS:2023geu} which provides a good fit to the data down to 100 eV. Below 100 eV we will smoothly extrapolate to 100\% quenching at 20 eV (a suggested lower-bound on charge creation is used in~\cite{Sarkis:2020soy}). In the absence of low-energy data for germanium we adopt the Sarkis model. The quenching factors are shown in Fig.~\ref{fig:quenching}. 

\begin{figure*}[t]
    \centering
    \includegraphics[width=0.45\textwidth]{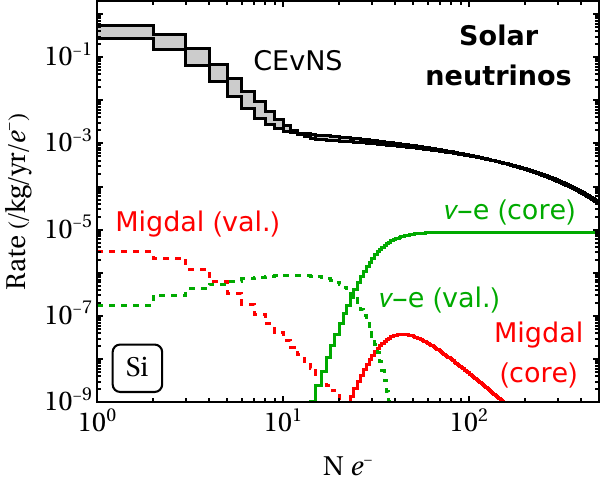}\hspace{5mm} 
    \includegraphics[width=0.45\textwidth]{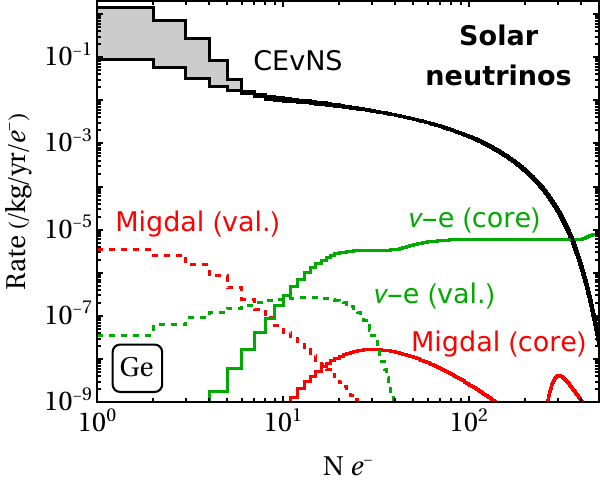}\\
    \caption{Neutrino-electron and neutrino-nucleus scattering rates in silicon (left) and germanium (right) in terms of liberated charge. The bands denote uncertainty due to the nuclear recoil  quenching model.}
    \label{fig:nu_iRates}
\end{figure*}

The differential rates of solar neutrinos scattering in germanium and silicon, for the various processes described in Sec.~\ref{sec:nuinducedelectronicexcitations}, are shown in Fig.~\ref{fig:nuRates}. The nuclear scattering is given in black, while the resulting Migdal signal is given in red, and electron scattering in green. The valence contribution (new result from this work) is shown as a dashed line and that of the core electrons is shown as a solid line. We can see that the valence contributions are relevant below 100 eV (40 eV) for silicon (germanium), but are substantially subdominant to the nuclear recoil rates. 

The observable signal measured by low-threshold solid-state detectors that we consider is the number of liberated electrons, $N_e$. We assume this is a statistical parameter which has some mean value, $\bar{N_e}$, determined only by the material, the energy deposited, and the type of scattering. When we are considering the ionization due to nuclear recoils we first include the effect of quenching (as discussed above). For an electron-equivalent energy deposit of $E_{ee}$ the mean number of ionized electrons is then
\begin{equation}
    \bar{N_e}(E_{ee}) = 1 + \lfloor \frac{E_{ee} - E_g}{\varepsilon}\rfloor,
\end{equation}
where $E_g$ and $\varepsilon$, band gap energy, and average energy to liberate an electron from the target, respectively. For silicon and germanium targets we take the values for these parameters to be:
\bea
E_g^{\rm{Si}} & = 1.11 \,\rm{eV}  \hspace{7mm} & \varepsilon^{\rm{Si}} = 3.6 \,\rm{eV} \nonumber\\
E_g^{\rm{Ge}} & = 0.67 \,\rm{eV}  \hspace{7mm} & \varepsilon^{\rm{Ge}} = 2.9 \,\rm{eV}\nonumber
\eea
The rate of ionization, $N_e$, is assumed to follow a Poisson distribution $p(N_i|N_e)$, and can be calculated by integrating over the deposited energy:
\begin{equation}
    \frac{d\ml{R}}{dN_e} = \int_0^{E^{\rm{max}}_R} dE \frac{d\ml{R}}{dE} p(\bar{N_e}(E_R)| N_e),
\end{equation}
where $E_R^{\rm{max}}$ is the maximum energy that can be deposited, which depends on the scattering target and incoming neutrino energy. In Fig.~\ref{fig:nu_iRates}  we show the calculated ionization rates for the same processes given in Fig.~\ref{fig:nuRates}. While measuring ionization alone does not permit discrimination, in the rest of the analysis we will consider two distinct scenarios, both with and without discrimination.

\section{Projected experimental reach and neutrino floors}
\label{sec:sensitivity}

We now turn to considering the potential reach of future germanium and silicon dark matter detectors. We will consider dark matter-electron scattering with light and heavy mediators, dark matter-nucleus scattering, and the Migdal effect. Additionally, we consider several beyond-Standard Model neutrino interactions, introduced in Sec.~\ref{sec:nuinducedelectronicexcitations}, where the neutrino also couples to either electrons or quarks via a new light-mediator. We project the experimental sensitivity of future germanium and silicon dark matter detectors to these light mediators through observations of solar neutrinos.

In the analysis that follows we are interested in the ultimate sensitivity of ionization detectors and so we choose to consider a threshold of one electron. Additionally, we restrict the region of interest to 1-400 ionized electrons to ensure the analysis remains computationally tractable. We note that this underestimates the sensitivity attainable at larger dark matter masses (where there is sufficient kinetic energy to ionize $>400$ electrons. However we expect that other experiments, e.g. xenon based experiments utilizing the Migdal effect, are better equipped to probe this region. Therefore, we will restrict the dark matter masses we consider to avoid painting a misleading picture in this region.

For the dark matter searches we assume that the dominant source of background is due to solar neutrinos, which scatter from both electrons and nuclei, the latter of which induces the Migdal effect.\footnote{Depending on the detector location one may also need to consider the flux of nearby nuclear reactors, which we do not consider here.} The CEvNS nuclear recoil background is much larger at small recoil energies, dominating over the electron scattering background over our entire region of interest (see Fig.~\ref{fig:nu_iRates}). Therefore, for the nuclear recoil search we will consider the CEvNS nuclear-recoil background only. However, for the electronic signals (electron scattering and the Migdal effect), we consider two background models: CEvNS nuclear recoils only and neutrino-electron scattering plus the Migdal effect from CEvNS. This corresponds to assumptions of no/perfect nuclear-electron recoil discrimination, respectively. For the beyond-Standard Model sensitivities, we assume there is no background.

\subsection{Statistical formalism}

We will define the discovery reach as the cross-section which produces a median $3\sigma$ deviation from the background only hypothesis. To compute these we will use the statistical formalism of~\cite{Cowan:2010js}, making use of the Asimov dataset and asymptotic formulae. The likelihood of a WIMP hypothesis (cross section $\sigma_\chi$ and mass $m_\chi$) is defined by:
\begin{equation}
    \mathcal{L}(\sigma_\chi,m_\chi|\mathbf{\theta}) = \prod_k f(\theta_k) \sum_i p\left(N^{\rm{obs}}_{e,i}|N^{\rm{exp}}_{e,i}(\sigma_\chi,m_\chi) \right)
\end{equation}
where the sum runs over ionization bins from $N=1$ to 400, and $p(N_i|N)$ is the Poisson probability of observing $N^{\rm{obs}}_{e,i}$ events given the expected $N^{\rm{exp}}_{e,i}(\sigma_\chi,m_\chi)$. The five nuisance parameters, $\mathbf{\theta}$, are the solar neutrino flux components: $pp$, $pep$, $^7$Be, $^{13}$N/$^{15}$O (which have a fixed flux ratio), and $^8$B. For the fluxes and their uncertainties we take the high-$Z$ model from~\cite{Haxton:2012wfz}. The uncertainties in the fluxes are assumed to be normally distributed, i.e. $f(\theta_i)$ is a normal distribution, centered at $\Phi_{\nu,i}$ and with standard deviation equal to the error $\Delta \Phi_{\nu,i}$.

The discovery significance of a particular hypothesis, $\{\sigma_\chi,m_\chi\}$, is calculated for the Asimov dataset (i.e. where $N^{\rm{obs}}_{e,i} = N^{\rm{exp}}_{e,i}(\sigma_\chi,m_\chi)$) through the test statistic:
\begin{equation}
    q_0(\sigma_\chi,m_\chi) = \begin{cases} -2 \log \frac{\mathcal{L}(\sigma_\chi = 0|\hat{\theta})}{\mathcal{L}(\hat{\sigma}_\chi,m_\chi|\hat{\hat{\theta}})} & \sigma_\chi > 0 \\
    0 & \sigma_\chi \leq 0 \\ \end{cases}
\end{equation}
where the hatted parameters maximize likelihoods. Through Wilks' theorem, the median significance can be found from $Z = \sqrt{q_0}$~\cite{Cowan:2010js}.

\begin{figure*}[t]
    \centering
    \includegraphics[height=5cm]{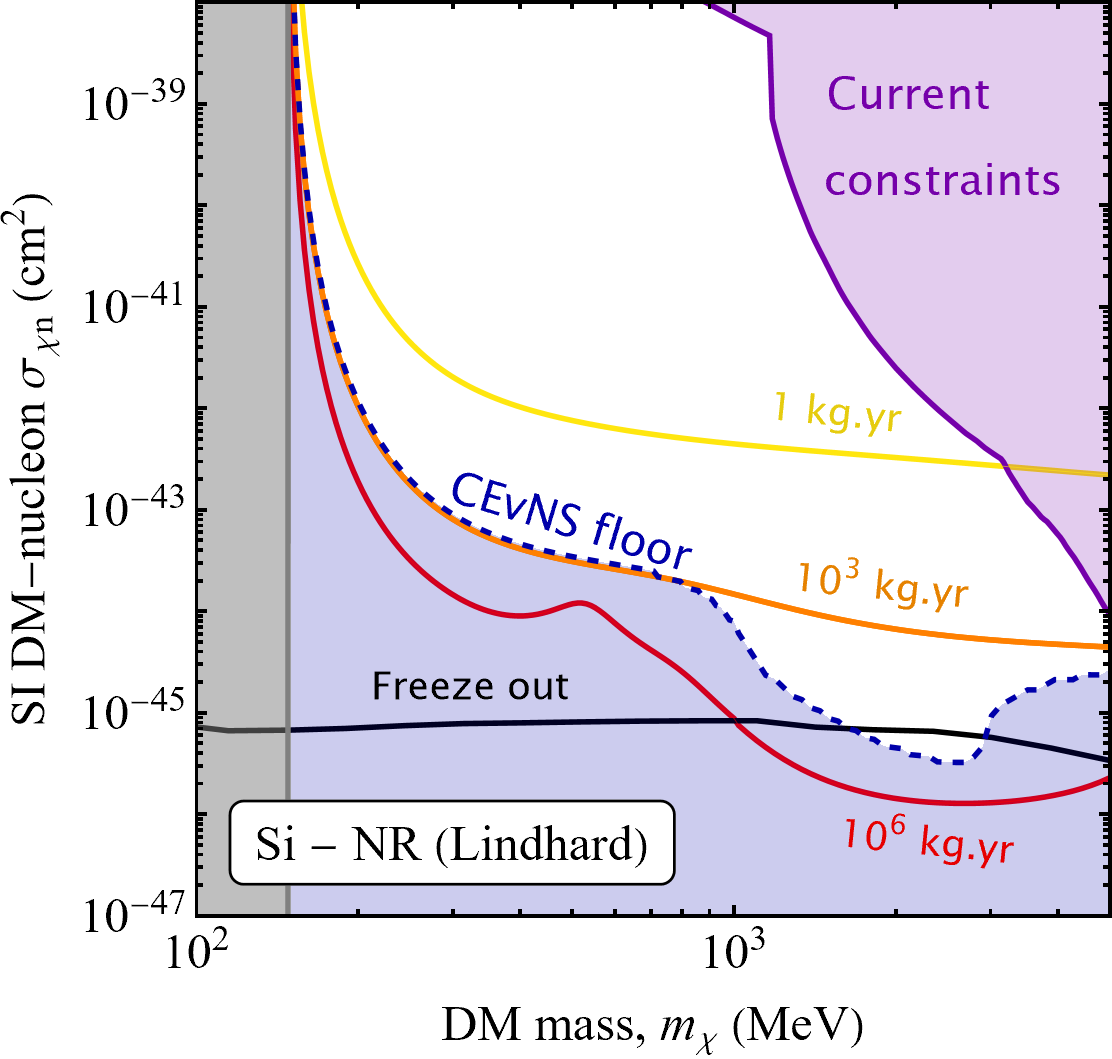} 
    \includegraphics[height=5cm]{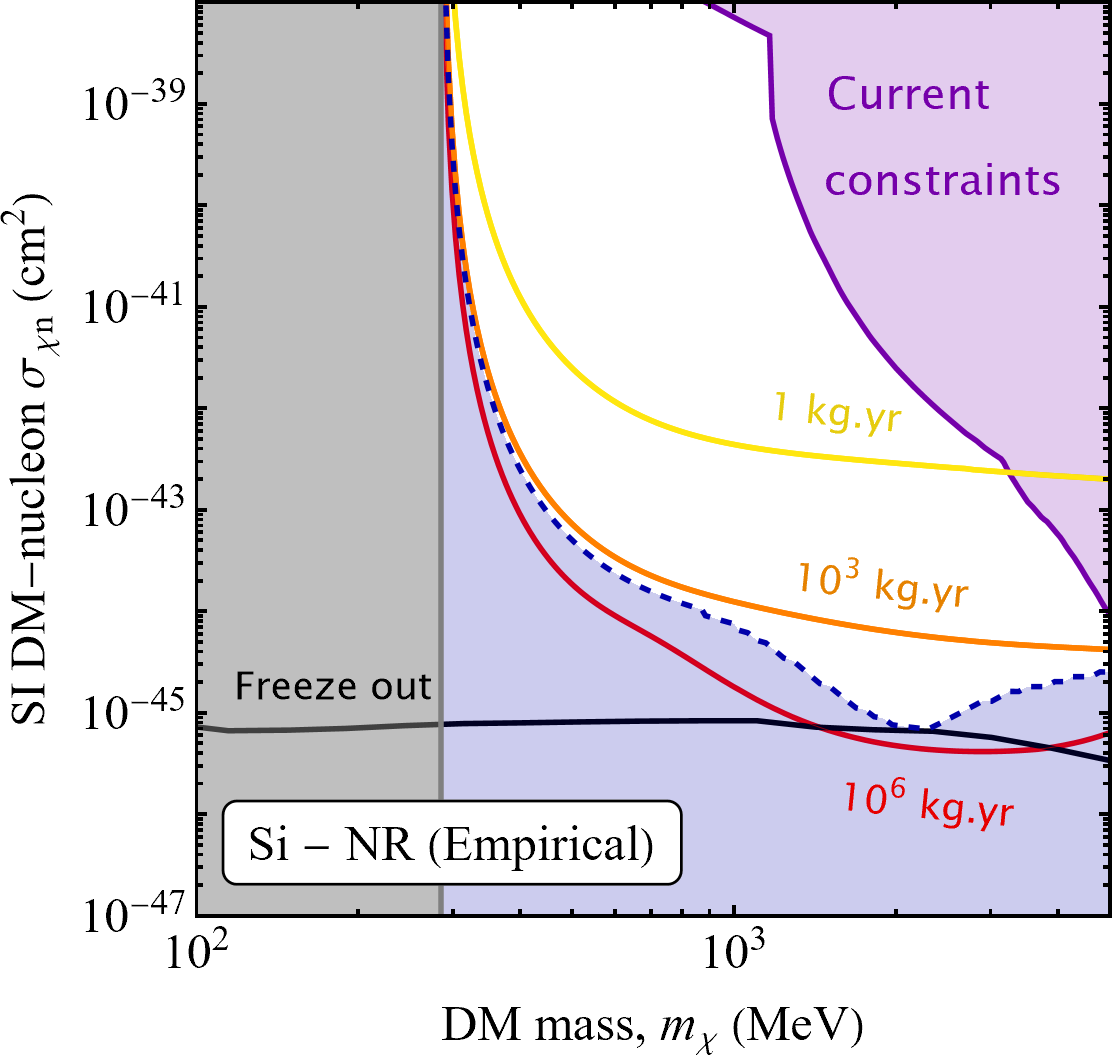}
    \includegraphics[height=5cm]{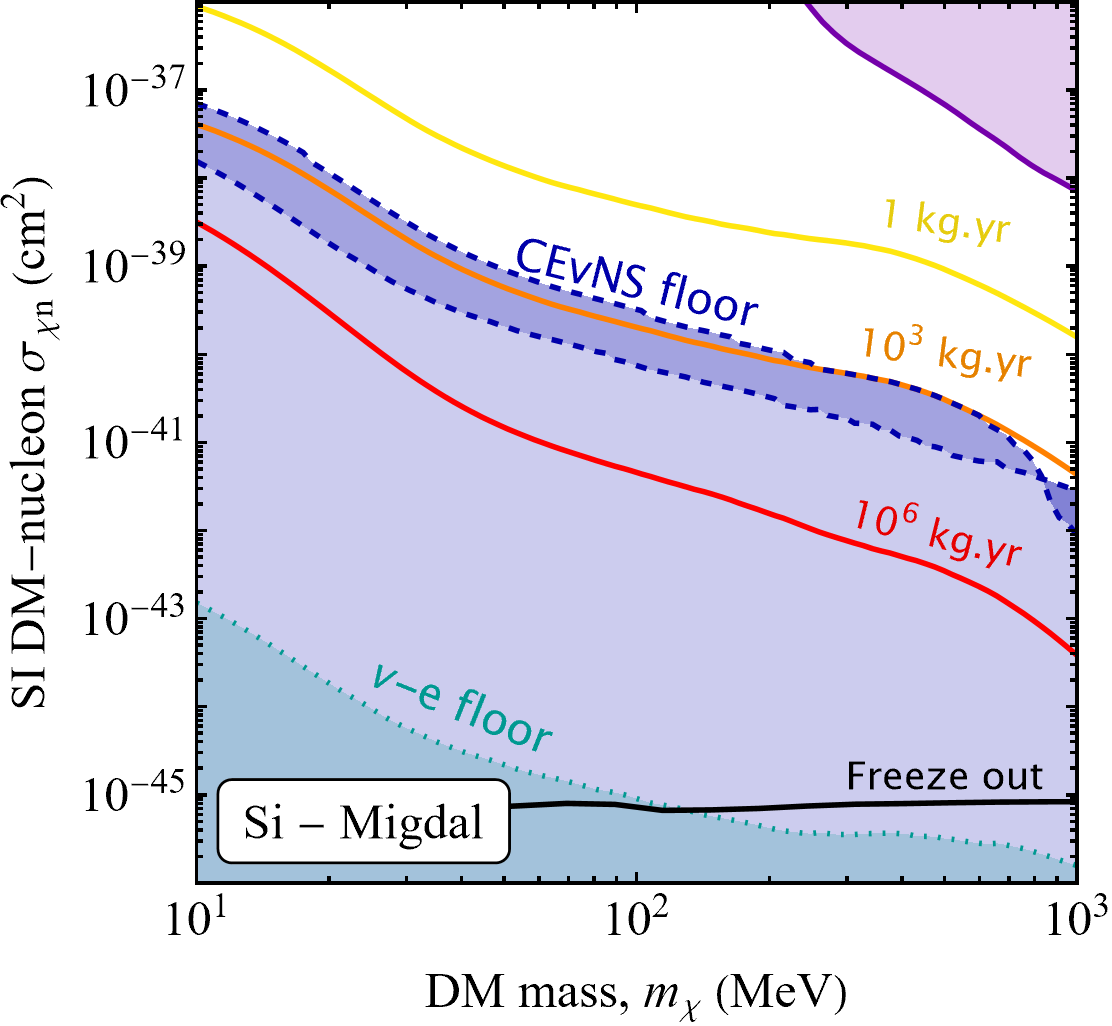}\\
    \includegraphics[height=5cm]{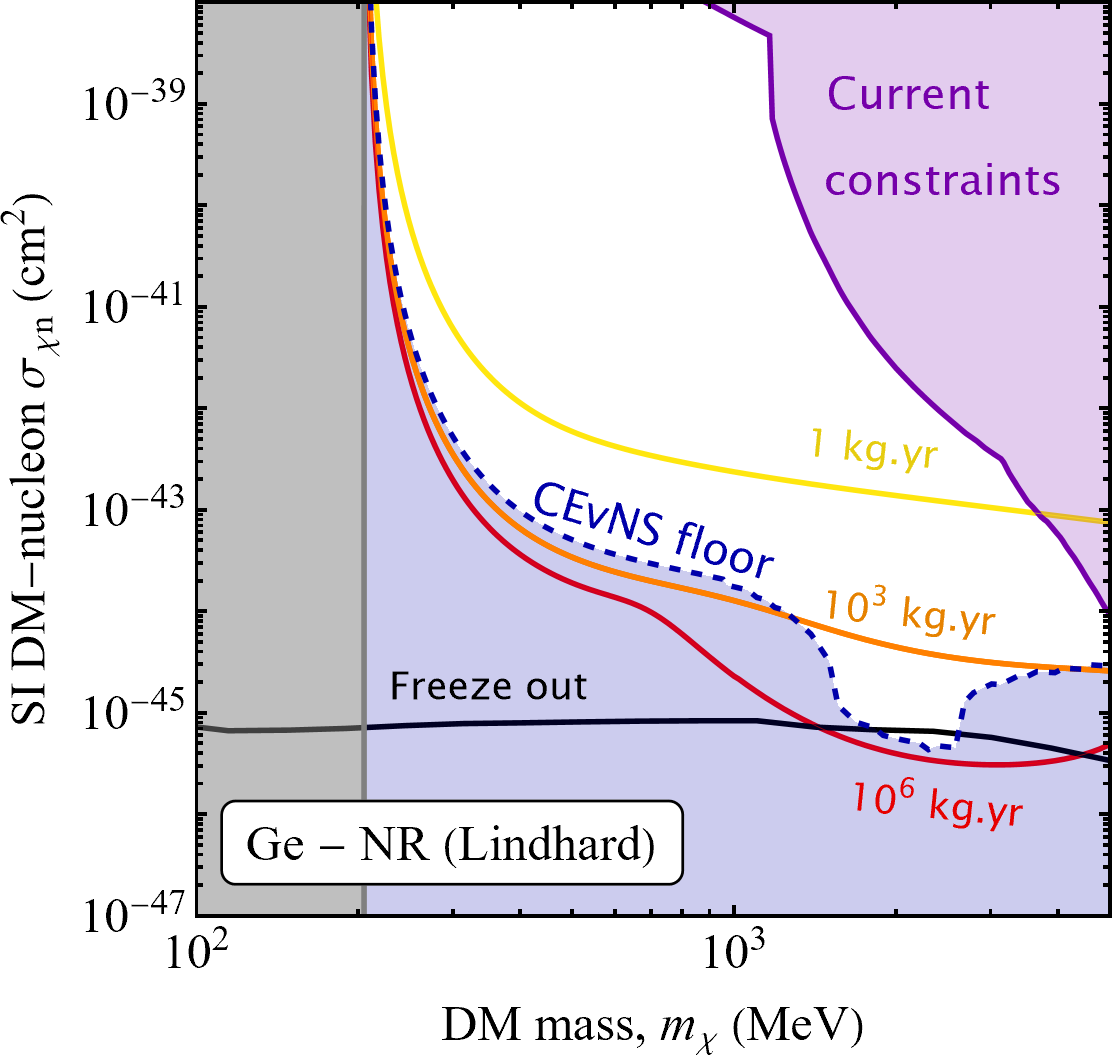} 
    \includegraphics[height=5cm]{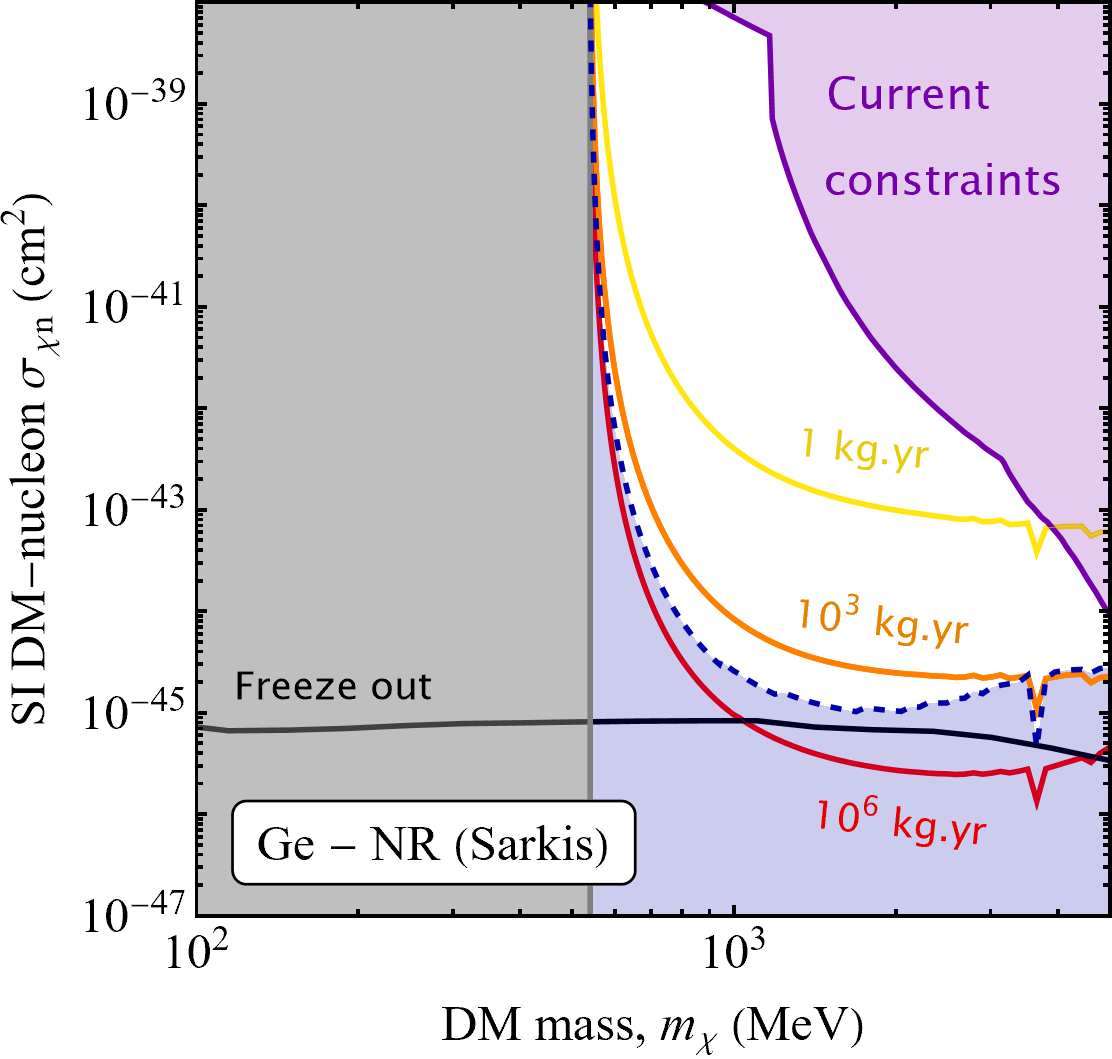}
    \includegraphics[height=5cm]{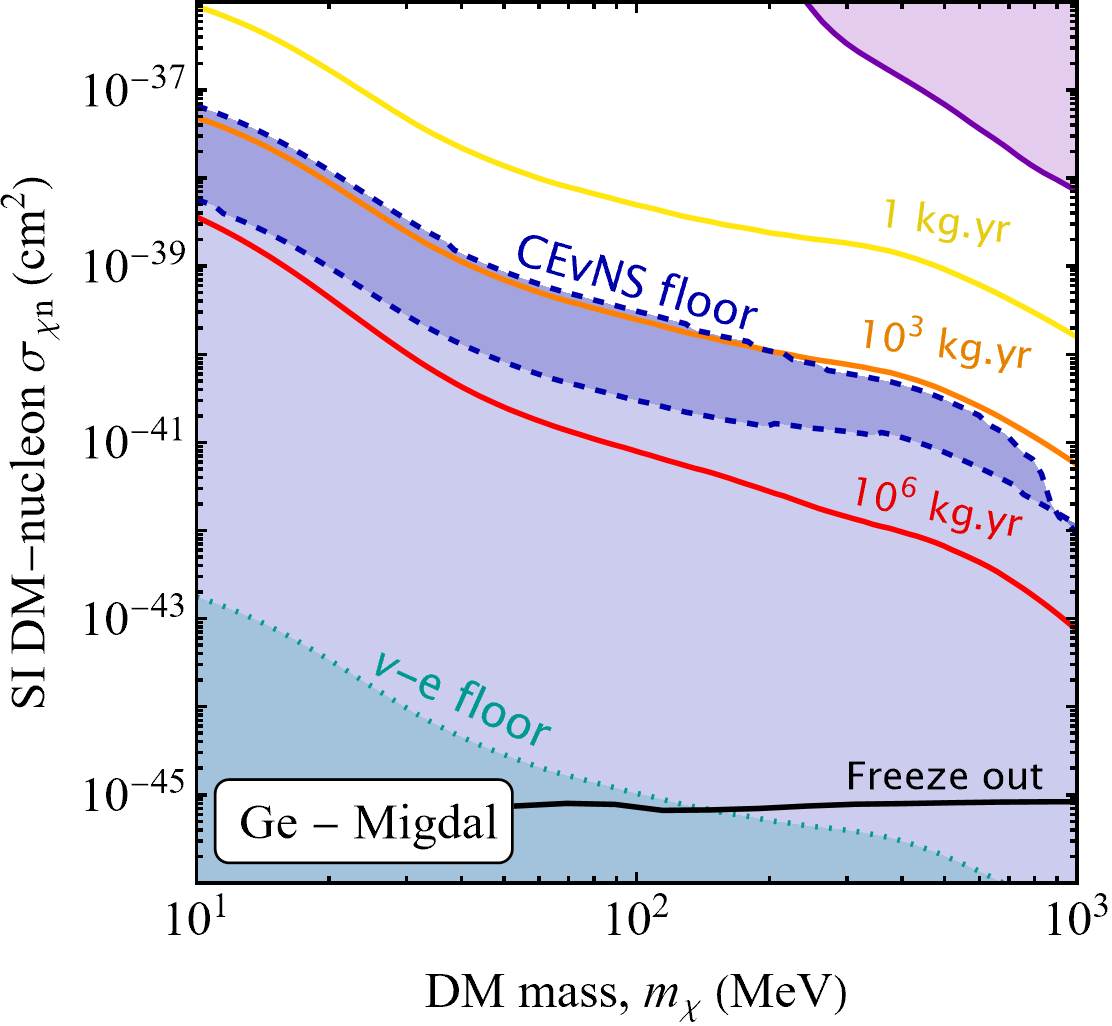}\\
    \caption{Silcon (top row) and germanium (bottom row) DM-nucleon discovery limits for NR -- Lindhard (left), NR -- alternate quenching model (center), and Migdal (right). The yellow, orange and red curves show the median 3$\sigma$ discovery limit for exposures of (1, 10$^3$, 10$^6$) kg-years, respectively. The black curve shows a thermal freeze-out theory benchmark for Majorana DM through a $Z'$-portal~\cite{Blanco:2019hah}. The blue dashed curve shows the neutrino floor due to CEvNS, while the teal dotted curve is the floor due to neutrino-electron scattering. The thickness of the CEvNS floor in the Migdal plot shows the effect of the two quenching models on the neutrino background, while the iso-exposure curves were calculated with the Lindhard model. The purple region in the upper-right of each plot shows the currently excluded region from a variety of experiments. The gray region on the left is inaccessible due to the single-electron ionization threshold.}
    \label{fig:NRdiscLim}
\end{figure*}

When the expected background from neutrinos is small compared to the dark matter signal, the $\sigma_\chi$ discovery limit evolves with increasing exposure ($MT$, i.e. the product of detector mass and time) according to $\sigma_\chi\propto 1/(MT)$. It is customary to define the figure of merit for how sensitivity increases with exposure as $n=-\left(\frac{d\sigma_\chi}{d MT}\right)^{-1}$. Therefore, for experiments initially in a background free regime $n=1$, but with increasingly larger exposures with non-zero backgrounds, one has $n>1$. 

In scenarios where the signal and backgrounds have close to degenerate spectra and there is a systematic uncertainty associated with the background normalization, $n$ can become large -- effectively halting the gain in sensitivity with increasing exposure. This phenomena is known as the neutrino floor~\cite{Billard:2013qya}. More generally, the region with $n>1$ is known as the neutrino fog~\cite{OHare:2021utq}. In this work we will use $n=2$ as the point of significant departure from background-free scaling and refer to it as the neutrino floor. 

This and related definitions of the neutrino floor are useful for gauging the parameter space that is ultimately able to be probed by a direct detection experiment. However, one must note that by this definition, the neutrino floor for each mass point is reached at a different exposure. Therefore the neutrino floor curve cannot correspond to any single experimental result. For this reason we will additionally calculate iso-exposure discovery limits for three representative exposures: 1, $10^3$ and $10^6$ kg-years.

\subsection{Dark matter-nucleon sensitivity}

First, we calculate the neutrino floor and discovery limits for the dark matter-nucleon cross section via nuclear scattering and the Migdal effect. We consider both silicon and germanium targets and the two quenching models introduced in Section~\ref{sec:results}: Lindhard and an empirical fit for silicon, and Lindhard and Sarkis for germanium). The results are shown in Fig.~\ref{fig:NRdiscLim}. As mentioned, given that CEvNS dominates over neutrino-electron scattering, we only have a CEvNS floor for the nuclear recoil case. For the Migdal case, however, we calculate two floors: a CEvNS floor (no discrimination) and an electron-neutrino floor (with discrimination). 

In the high-mass region we find that silicon and germanium have very similar discovery limits. However, in the low-mass region silicon has a significant edge due to its lighter mass -- despite its larger band-gap. Additionally, when considering the alternate quenching models, which are more pessimistic than the Lindhard model at low-recoil energies, the minimum mass of the kinematically allowed region is greatly increased. This is particularly pronounced for the germanium case, where the Sarkis model predicts no ionization below $E_R = 40$ eV, leading to a minimum mass of $m_\chi\sim55$ MeV. However, this region can still be probed via the Migdal effect for cross sections down to $\sim10^{-40}$ cm$^2$.

As observed by numerous previous studies~\cite{Billard:2013qya,Ruppin:2014bra,OHare:2016pjy,Dent:2016iht,Dent:2016wcr,Ng:2017aur,AristizabalSierra:2017joc,Boehm:2018sux,Carew:2023qrj}, we find that in the background free region, the discovery limit initially evolves $\propto$ exposure$^{-1}$. Then, except at particular masses where the dark matter recoil spectrum mimics that of certain neutrino species, it evolves $\propto$ exposure$^{-1/2}$. However, in the present work, we are considering the smeared ionization spectrum, and not the direct nuclear recoil spectrum. This has the effect of smearing the low-energy neutrino flux components together, leading to a situation where we do not observe degeneracies with each flux component at correspondingly different dark matter masses. 

In the case of germanium, the neutrino background is equally degenerate with dark matter masses below $m_\chi\sim60$ MeV. While in the case of silicon, we see a single mass point with a pronounced degeneracy at $m_\chi\sim50$ MeV. For both silicon and germanium we see a dip in the neutrino floor above $m_\chi>1$ GeV. This is due to the ionization spectra being quite different from the low-energy solar flux components and the ${}^8\textnormal{B}$ component. For masses more than a few GeV the dark matter spectra start to become degenerate with the ${}^8\textnormal{B}$ component, producing the effect that the floor moves to larger cross sections. 

For the direct nuclear recoil search, the neutrino floor sits well above thermal freeze-out benchmark, except for a small region around 1 GeV. However, the exposure required to have sensitivities reaching the benchmark is of the order $10^6$ kg-years -- infeasibly large for this type of detector. Encouragingly, more reasonable exposures of $10^3$ kg-years can attain sensitivities comparable to the neutrino floor for light masses. 

For the Migdal search, the thermal freeze-out benchmark is 5-6 orders of magnitude below the CEvNS floor, yet above the neutrino-electron floor for masses above 100 MeV. To reach the freeze-out benchmark (via the Migdal effect), a detector would not only need the capability to discriminate and remove the CEvNS nuclear recoil background to one part per million, but also require extremely large exposures of order $10^{9}$ kg-years.

The Migdal channel is expected to have superior sensitivity to the nuclear recoil channel below around $m_\chi = 150 (220)$ MeV for silicon (germanium) with the Lindhard quenching model and $m_\chi = 300(550)$ MeV for silicon (germanium) with the alternate quenching model.

Somewhat counterintuitively, should the more pessimistic quenching models reflect reality, the experimental sensitivity does not suffer greatly. There are two reasons for this: first, the more significant quenching will also apply to the background, suppressing it, and second, at very low masses where sensitivity falls off steeply, the Migdal effect becomes the more sensitive search method, and its discovery reach is only improved by the background suppression due to quenching.

\subsection{Dark matter-electron sensitivity}

\begin{figure*}[t]
    \centering
    \includegraphics[width=0.45\textwidth]{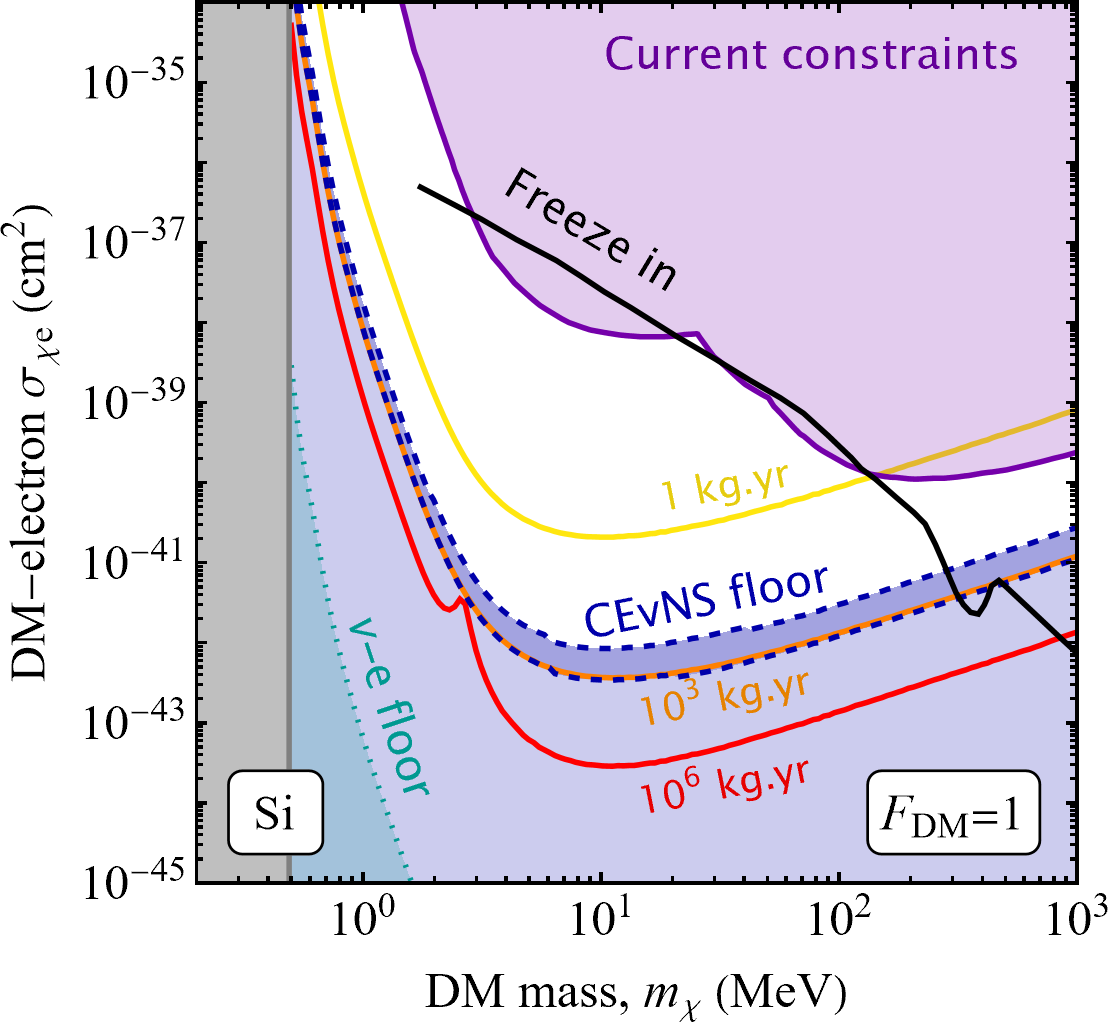}\hspace{5mm} 
    \includegraphics[width=0.45\textwidth]{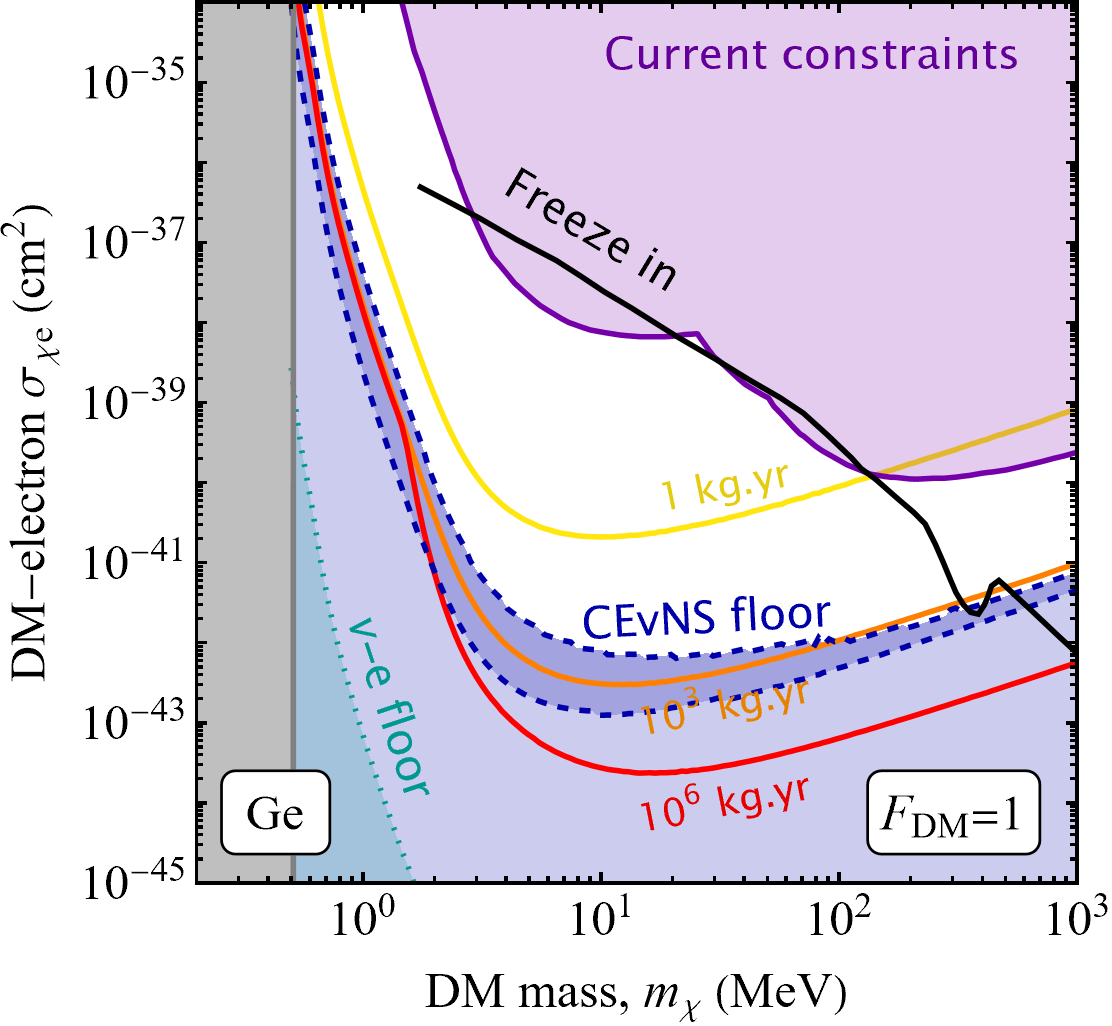}\\
    \includegraphics[width=0.45\textwidth]{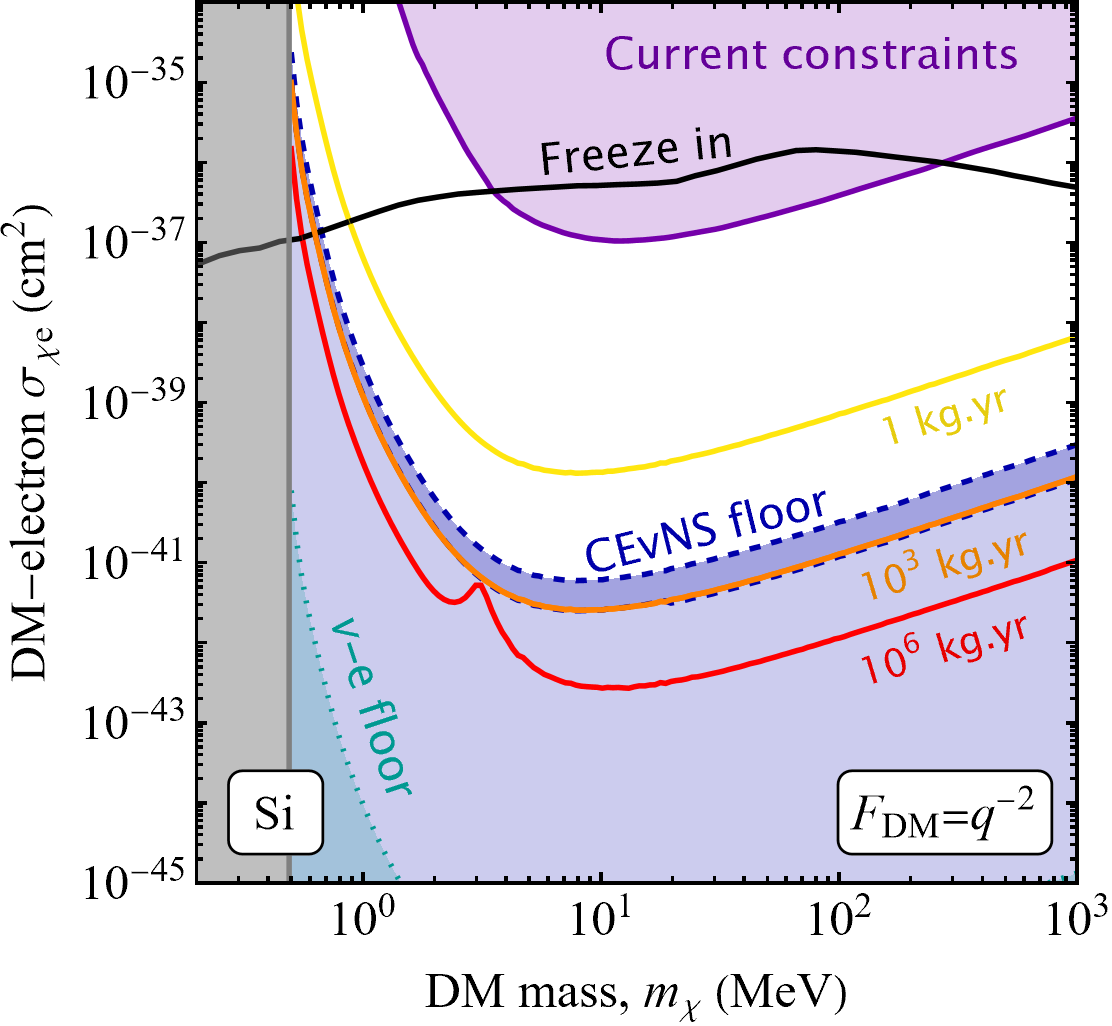}\hspace{5mm} 
    \includegraphics[width=0.45\textwidth]{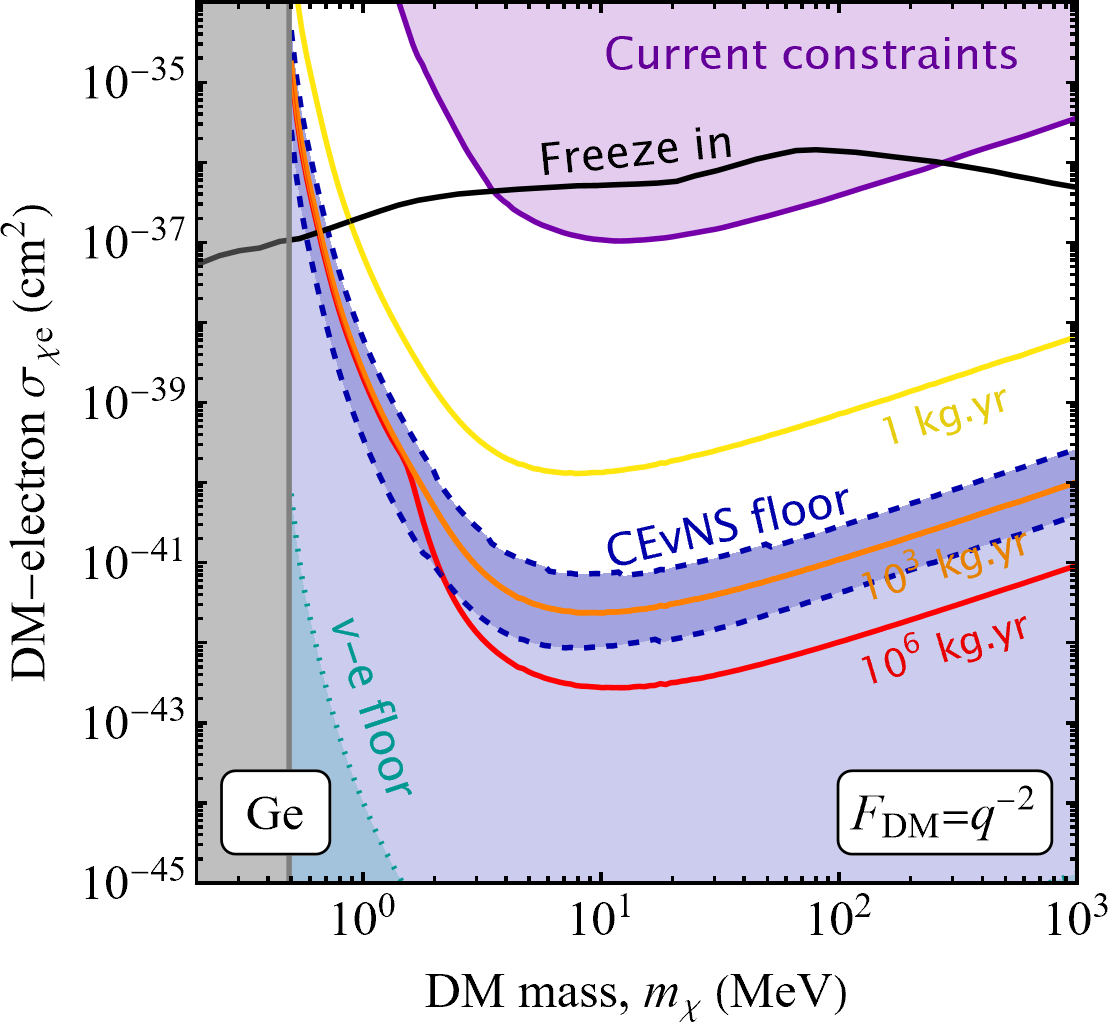}\\
    \caption{DM-electron discovery limits for silicon (left column), germanium (right column), assuming heavy (top row) and light (bottom row) mediators. The yellow, orange and red curves show the median 3$\sigma$ discovery limit for exposures of (1, 10$^3$, 10$^6$) kg-years, respectively, under the CEvNS background and assume Lindhard quenching. The blue dashed curve shows the CEvNS floor, which the thickness showing the effect of the two quenching models. The black curve shows a freeze-in theory benchmark~\cite{Essig:2022dfa}. The purple region in the upper and upper-right areas shows the excluded region from a variety of experiments. The gray region on the left is inaccessible due to the single-electron ionization threshold.}
    \label{fig:EdiscLim}
\end{figure*}

For dark matter scattering on electrons we consider scattering both via light and heavy mediators, and compute neutrino floors for both the discrimination and non-discrimination scenarios. The results of these, along with iso-exposure discovery limits are shown in the panels of Fig.~\ref{fig:EdiscLim}. As before (in the Migdal case) we found that the neutrino-electron floor is $5\textendash6$ orders of magnitude below the CEvNS floor, which is about what we should expect given the relative difference in the scattering rates. However, as before, this isn't good news as it requires a proportionally larger exposure, of order $10^{9}$ kg-years, to reach it. Such large exposures are infeasible for this detector technology. Therefore we don't show the full neutrino-electron floor on the plots, but include it's indicative location for completeness.

In general the discovery potential is about the same for silicon and germanium targets, which germanium being slightly better at larger masses while suffering for a degeneracy at lighter masses. The degeneracy is evident from the lack of improvement in germanium's discovery limit going from 10$^3$ to 10$^6$ kg-years. This is similar to the effect seen for the nuclear recoil search where the CEvNS rates from the low-energy solar components have smeared together. Silicon, on the other hand, exhibits a degeneracy at a particular mass: $m_\chi \sim 2.8$ MeV for heavy-mediator scattering, and $m_\chi \sim 3.0$ MeV for light-mediator scattering. Interestingly, these degeneracies exist between the ionization spectra of $^7$Be CEvNS and dark matter-electron scattering.

As with the nuclear recoil sensitivity, an exposure of 10$^3$ kg-years is sufficient to reach the floor in all cases and across all masses. Such an exposure would probe the freeze-in benchmark across a very wide range of masses, only missing it below an MeV for light mediators and above a few hundred MeV for the heavy mediator case.

\subsection{Neutrino BSM sensitivity}

\begin{figure*}[t]
    \centering
    \includegraphics[width=5.5cm]{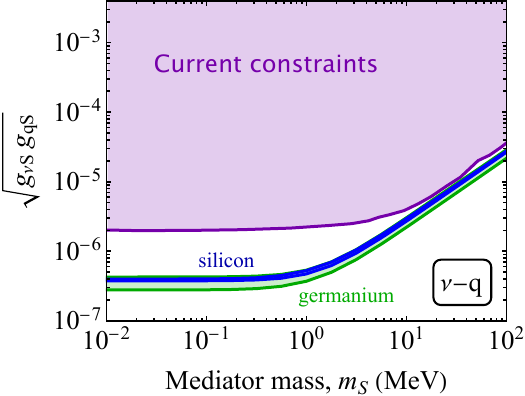}\hspace{-2mm} 
    \includegraphics[width=5.5cm]{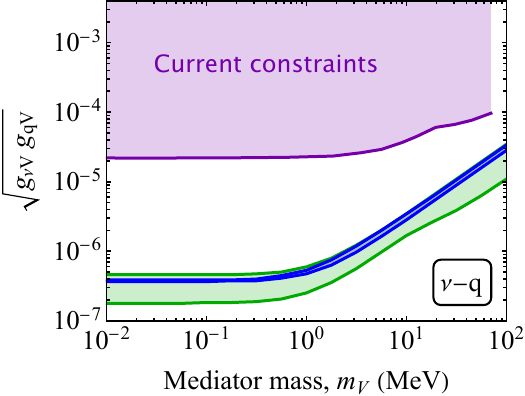}\hspace{-2mm} 
    \includegraphics[width=5.5cm]{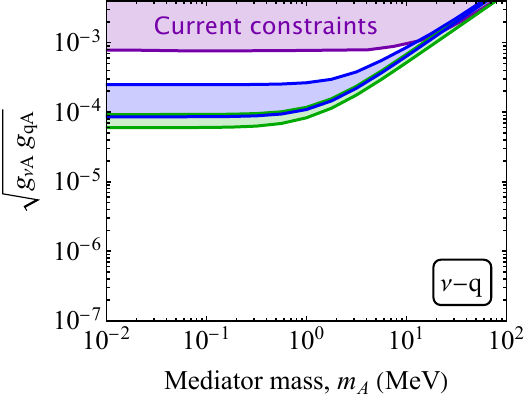} \\
    \caption{Discovery limits for a $10^3$ tonne-year exposure for neutrino-quark BSM: scalar (left), vector (middle), and axial-vector (right) mediators. The shaded blue and green bands represent the uncertainty due to the quenching factor for silicon and germanium respectively. Prior constraints are taken from~\cite{Schwemberger:2022fjl}.}
    \label{fig:nuNdiscLim}
\end{figure*}

Now we turn to the consideration of the sensitivity that a dark matter experiment will have to beyond Standard Model physics in the neutrino sector. By virtue of the preceding analyses, we found that an exposure of 10$^3$ kg-years is sufficient to reach the CEvNS neutrino floor -- thus furnishing a significant number of solar neutrino events. While the motivation of these experiments is to search for dark matter, such a dataset would be ripe for further analysis, e.g. as tests of the Standard Model. Other large dark matter experiments will also probe this parameter space. However, low-threshold detectors will have a distinct advantage when, for example, light mediators modify the low-energy recoil spectrum.

\begin{figure*}[t]
    \centering
    \includegraphics[width=0.45\textwidth]{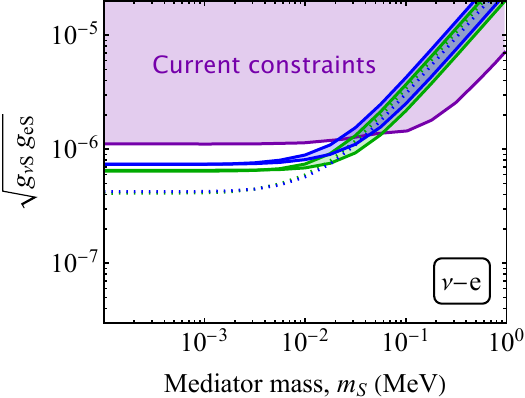}\hspace{5mm} 
    \includegraphics[width=0.45\textwidth]{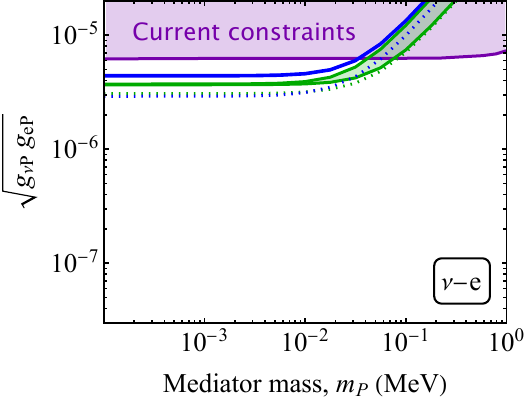}\\
    \includegraphics[width=0.45\textwidth]{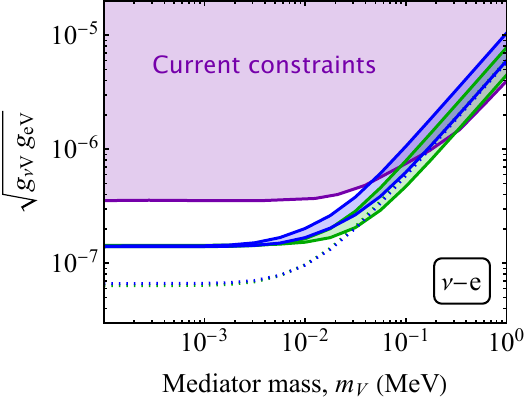}\hspace{5mm} 
    \includegraphics[width=0.45\textwidth]{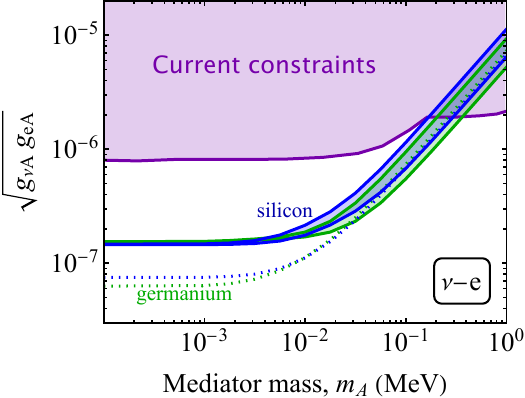}\\
    \caption{Discovery limits with a $10^3$ tonne-year exposure for neutrino-electron BSM: scalar (top left), pseudo-scalar (top right), vector (bottom left) and axial-vector (bottom right) mediators.  The shaded blue and green bands represent the uncertainty due to the quenching factor (which affects the CEvNS background model) for silicon and germanium, respectively. The dotted curve shows the discovery limits without the CEvNS background. Prior constraints taken from~\cite{Schwemberger:2022fjl}.}
    \label{fig:nuEdiscLim}
\end{figure*}

To demonstrate this, we compute the discovery potential for a single 10$^3$ kg-years exposure with a 1-electron threshold and assume no other sources of background beyond neutrino scattering. These are optimistic assumptions and thus we consider our results a best-case scenario for the sensitivity this setup can achieve. We include three scenarios: Lindhard quenching, our alternate quenching, and nuclear recoil discrimination. The discrimination scenario only benefits the searches for BSM physics in the electron scattering channel, shown in Fig.~\ref{fig:nuEdiscLim}, and so the nuclear recoil analyses only have two scenarios, shown in Fig.~\ref{fig:nuNdiscLim}.  

For consistency with previous work we show the sensitivity with respect to the root of the product of the neutrino-mediator and quark/electron-mediator coupling. In Figs.~\ref{fig:nuNdiscLim} and~\ref{fig:nuEdiscLim} we have shown the prior constraints on this combination of couplings from neutrino scattering, taken from~\cite{Schwemberger:2022fjl}. However, we note that there will be constraints from other sources (e.g. meson decay) on the quark/electron-mediator coupling. With the visible coupling driven to smaller values, the neutrino-mediator coupling constraint will be correspondingly weaker.

The sensitivity to new light scalar and axial vector mediators through the nuclear recoil channel is found to improve over prior constraints by around an order of magnitude. The vector mediator case is improved by two orders of magnitude owing to the interference effects, which makes sensitivity scale proportional to the product of the couplings (rather than couplings squared). We find that the germanium sensitivity is consistently slightly stronger than the silicon sensitivity. For the scalar and vector mediators this is because they induce a coherent enhancement which advantages germanium's larger nucleus. While for the axial-vector mediator, germanium is advantaged because it contains a larger fraction of odd-nuclei (compared to silicon), which carry net spin and are thus able to couple to axial currents. Lastly, germanium has a smaller bandgap, which gives it an advantage near threshold, and a lower average energy to excite an electron, which helps it statistically.

Turning now to the electron coupling case, we see a similar pattern to the nucleon results. The projected sensitivity is better than previous constraints, but only by a factor of $\sim 2$ or less, and only at very small mediator masses (i.e. below 100 keV). This is a smaller relative improvement with respect to previous constraints compared to the nucleon case. This is because, for the electron channel, the previous constraints come from XENON1T, which has a comparable exposure but higher threshold and background than we consider. 

In every case germanium is slightly more sensitive than silicon, given the lack of coherent enhancement in neutrino-electron scattering, this is likely due to germanium's smaller bandgap and lower average excitation energy. When the CEvNS background is removed the sensitivity becomes a factor of $\lesssim 2$ better. This apparently small improvement is due to the fact the BSM rate goes as the fourth power of the $\sqrt{g_\nu g_e}$ parameter.

\section{Summary}
\label{sec:summary}

In this work we completed a comprehensive study of the dominant interaction channels of solar neutrinos with solid-state materials at very low energies. This involves the neutral current scattering of neutrinos from both the electrons and nuclei in the solid. The novelty of this work lies in treating the collective electronic excitations that can occur due to  interactions with the valence electrons and also from nuclear scattering (via a Migdal-like effect). We found that these excitations enhance the expected rate of neutrino scattering at very low energies. 

To treat the collective excitations we employed the Foldy-Wouthuysen transformation and then calculated the scattering rates following the Kramers-Heisenberg approach. This allowed us to express the neutrino-electron scattering rate in terms of five distinct structure factors, which depend on the properties of the solid. The leading structure factor was identified with the electron energy-loss function, while the others relate to magnetic properties of the solid (the latter being left to study in a future work.)

Motivated by the existence of low-threshold dark matter experiments which are exquisitely sensitive to ionization, we then projected the sensitivities they can achieve to both dark matter and BSM neutrino interactions. The neutrino scattering rates we calculated for two dielectric materials: silicon and germanium, making use of electron loss functions computed from density functional theory. These rates then served as both a signal and a background in the sensitivity calculations. 

We computed the ultimate sensitivity to dark matter that such ionization detectors can achieve for several exposures, as well as the corresponding neutrino-floors. We found that the silicon and germanium detectors perform very similarly, with silicon being slightly better at lower masses for the nuclear recoil channel. The neutrino floor can be reached for a large range of masses with an exposure of around $10^3$ kg-years. For the nuclear channel, the accessible parameter space lies almost entirely above the thermal freeze out benchmark and the Migdal effect gives the dominant sensitivity below around $m_\chi = 150 (220)$ MeV for silicon (germanium). While for the electron channel, $10^3$ kg-years can reach the floor across all the considered masses (from $< 1$ MeV to over 1 GeV), and probing the freeze-in scenario across almost this entire range for both light and heavy mediator couplings.

These low-threshold detectors, while searching for dark matter, will also be sensitive to departures from the Standard Model through their measurement of the solar-neutrino scattering rate. Using a set of simple, light-mediator models, we show that, at the exposures needed to reach the neutrino floor, these detectors would be sensitive to previously unconstrained regions of parameter space.

\section{Acknowledgments}
We thank Robert Calkins, Tongyan Lin, and Yonatan Kahn for useful discussions. The work of JLN is supported by the Australian Research Council through the ARC center of Excellence for Dark Matter Particle Physics, CE200100008. JBD acknowledges support from the National Science Foundation under grant
no. PHY-2412995. JBD thanks the Mitchell Institute at Texas A\&M University for its hospitality where part
of this work was completed.

\appendix
\section{Gamma matrix conventions}
\label{app:conventions}
The gamma matrices as well as the $\vec{\alpha}$ and $\beta$ Dirac matrices are taken to be in the Dirac representation:
\bea
\alpha_i=\begin{pmatrix}
0 & \sigma_i \\
\sigma_i & 0 
\end{pmatrix},\
\beta = \begin{pmatrix}
I & 0\\
0 & -I
\end{pmatrix},
 \gamma^0=\begin{pmatrix}
I & 0\\
0 & -I
\end{pmatrix},\
\gamma^i=\begin{pmatrix}
0 & \sigma_i\\
-\sigma_i & 0
\end{pmatrix},
\gamma^5=\begin{pmatrix}
0 & I\\
I & 0
\end{pmatrix},
\eea
and we are using the standard mostly minus metric.

\section{Electron-Neutrino Interactions}
\label{app:enuinteractions}
\subsection{Interaction Lagrangian} From the Lagrangian density in Eq.~(\ref{eq:lagrangian}), we extract the interaction Hamiltonian operator,
 \bea\label{e_nu_op}
 {\ml{H}}_{int}=\frac{G_F}{\sqrt{2}}\left[V_\mu-A_\mu\right]\left[(g_R+g_L)\ \gamma^0\gamma^\mu+(g_R-g_L)\ \gamma^0\gamma^\mu\gamma^5\right]
 \eea
where $V_\mu$ and $A_\mu$ are the neutrino vector and axial currents respectively. We add in the free Dirac Hamiltonian (in addition there is also the free neutrino field term which we suppress below), so the total electronic Hamiltonian becomes,
\bea
\hat{\ml{H}}_{e^{-}} &=& \vec{\alpha}\cdot\vec{p}+\beta m_e+ \frac{G_F}{\sqrt{2}}\ \left[V_\mu-A_\mu\right]\left[(g_R+g_L)\ \gamma^0\gamma^\mu+(g_R-g_L)\ \gamma^0\gamma^\mu\gamma^5\right]\nn\\
&=&\vec{\alpha}\cdot\vec{p}+\beta m_e+ \frac{G_F}{\sqrt{2}}\ \left[V^0-A^0\right]\ \left[(g_R+g_L)+(g_R-g_L)\ \gamma^5\right]\nn\\
&&- \frac{G_F}{\sqrt{2}}\ \left[\vec{V}-\vec{A}\right]\cdot\vec{\alpha} \left[(g_R+g_L)+(g_R-g_L)\ \gamma^5\right]\nn\\
&=& \beta m_e + \ml{O}_\alpha+\ml{O}_5+\ml{O}_{\alpha 5}+\ml{E}_{VA},
\eea
where we have defined,
\bea
\ml{O}_\alpha &\equiv& \vec{\alpha}\cdot\left[\vec{p}-\frac{G_F}{\sqrt{2}}\ \left[\vec{V}-\vec{A}\right] \left(g_R+g_L\right)\right]\nn\\
\ml{O}_5 &\equiv& \frac{G_F}{\sqrt{2}}\ (g_R-g_L)\ \gamma^5\left[V^0-A^0\right]\nn\\
\ml{O}_{\alpha5} &\equiv& - \frac{G_F}{\sqrt{2}}\ (g_R-g_L)\ \gamma^5 \vec{\alpha}\cdot\left[\vec{V}-\vec{A}\right]\nn\\
\ml{E}_{VA} &\equiv& \frac{G_F}{\sqrt{2}}\ (g_R+g_L)\left[V^0-A^0\right]
\eea
These operators satisfy the following product rules, 
\bea
\beta\ml{O}_\alpha &=& -\ml{O}_\alpha\beta\;\;;\;\;
\beta\ml{O}_5 = -\ml{O}_5\beta\;\;;\;\;
\beta\ml{O}_{\alpha5} = \ml{O}_{\alpha5}\beta\;\;;\;\;
\beta\ml{E}_{VA} = \ml{E}_{VA}\beta
\eea
Defining
\bea
\ml{E} \equiv \ml{E}_{VA}+\ml{O}_{\alpha5}\nn\;\;;\;\;
\ml{O} \equiv \ml{O}_\alpha+\ml{O}_5
\eea
we find that the electronic Hamiltonian operator is 
\bea
\hat{\ml{H}}_{e^{-}} &=& \beta m_e +\ml{O}+\ml{E}
\eea
where $\ml{O}$ is odd and $\ml{E}$ is even with respect to the rest mass operator $\beta m_e$,
\bea
\left(\beta m_e\right)\ml{O} = -\ml{O}\left(\beta m_e\right)\nn\;\;;\;\;
\left(\beta m_e\right)\ml{E} = \ml{E}\left(\beta m_e\right)
\eea

\subsection{Foldy–Wouthuysen transformation}
\label{app:foldy}

At this point, we proceed as in Chapter~$4$ of \cite{Bjorken:100769} by applying a Foldy-Wouthuysen transformation~\cite{PhysRev.78.29}, whose purpose is to remove odd operators from the equation of motion. This is done by introducing a unitary transformation $U_{FW} = \textnormal{exp}\left(iS\right)$, where $S$ is Hermitian. Acting on the equation of motion one gets
\begin{align}
    e^{iS}\ml{\hat{H}}\psi = e^{iS}\ml{\hat{H}}e^{-iS}\psi' = \ml{\hat{H}}'\psi'
\end{align}
where $\ml{\hat{H}}$ will be left with no odd operators.

In our case we expand $e^{iS}\hat{\ml{H}}_{e^{-}}e^{-iS}$ in powers of inverse electron rest-mass energy, $1/m_e$, which is larger than the typical energy transfer we are considering. The result up to $O(1/m_e^{3})$ is
\bea\label{fw}
\hat{\ml{H}}^{(3)}_{e^{-}} = \beta\left(m_e+\frac{\ml{O}^2}{2m_e}-\frac{\ml{O}^4}{8m_e^3}\right)+\ml{E}-\frac{1}{8m_e^2}\left[\ml{O},\left[\ml{O},\ml{E}\right]\right]-\frac{i}{8m_e^2}\left[\ml{O},\dot{\ml{O}}\right]\nn\\
\eea
where $\dot{\ml{O}}$ is the time derivative of $\ml{O}$. The first term is,
\bea
\frac{\ml{O}^2}{2m_e}=\frac{(\ml{O}_\alpha+\ml{O}_5)^2}{2m_e}
\eea
where
\bea
\ml{O}_\alpha +\ml{O}_5 &=&  \vec{\alpha}\cdot\left[\vec{p}-\frac{G_F}{\sqrt{2}}\ \left[\vec{V}-\vec{A}\right] \left(g_R+g_L\right)\right]+\frac{G_F}{\sqrt{2}}\ (g_R-g_L)\ \gamma^5\left[V^0-A^0\right]\nn\\
&\equiv&  \vec{\alpha}\cdot[\vec{p}-\vec{\ml{V}}]+ \gamma^5\ml{A}
\eea
and 
\bea
\ml{O}_\alpha^2&=& \left[\left(\vec{p}-{\vec{\ml{V}}}\right)^2-\vec{\sigma}\cdot\left(\vec{\nabla}\times\vec{\ml{V}}\right)\right]\nn\\
&\approx&\left[\vec{p}\cdot\vec{p}-2(\vec{\ml{V}}\cdot\vec{p})-(\vec{p}\cdot\vec{\ml{V}})-\vec{\sigma}\cdot\left(\vec{\nabla}\times\vec{\ml{V}}\right)\right]
\eea
where 
\bea
\vec{\ml{V}} &=& \frac{G_F}{\sqrt{2}}\ \left(g_R+g_L\right) \left[\vec{V}-\vec{A}\right]\nn\\
{\ml{A}} &=& \frac{G_F}{\sqrt{2}}\ (g_R-g_L)\ \left[V^0-A^0\right]
\eea
Note that since $G_F\sim O({1}/{m_e^2})$, $\ml{O}_5^2=\ml{A}^2\sim O(G_F^2)$, thus we neglect it below,
\bea
\ml{O}_5\ml{O}_\alpha+\ml{O}_\alpha\ml{O}_5 &=& \vec{\sigma}\cdot\left[2\ml{A}\ \vec{p}-i (\vec{\nabla}\ml{A})\textendash\left\{\ml{V},\ml{A}\right\}\right]\mathbbm{1}\nn\\
&\approx& \vec{\sigma}\cdot\left[2\ml{A}\ \vec{p}-i (\vec{\nabla}\ml{A})\right]
\eea
where we have also neglected the term $\left\{\ml{V},\ml{A}\right\}$ because it is also $O(G_F^2)$.\\
The third term is proportional to $\ml{O}^4$, which is
\bea
\ml{O}^4&=&(\ml{O}_\alpha+\ml{O}_5)^4=\left(\vec{\alpha}\cdot[\vec{p}-\vec{\ml{V}}]+ \gamma^5\ml{A}\right)^4\approx p^4
\eea
Since the pre-factor is $\sim 1/(m_e^3)$, even a term linear in $G_F$ above would combine with the $\ml{O}^4$ term to give expressions of $O(1/m_e^{5})$ or smaller.\\

Next, the even (with respect to
$\beta$) term $\ml{E}$ is relatively straightforward,
\bea
\ml{E} &=&  \frac{G_F}{\sqrt{2}}\left[(g_R+g_L)\ [V^0-A^0]-(g_R-g_L)\ \vec{\sigma}\cdot[\vec{V}-\vec{A}]\right]\nn\\
&=&\left[\left(\frac{g_R+g_L}{g_R-g_L}\right)\ml{A}-\left(\frac{g_R-g_L}{g_R+g_L}\right)(\vec{\sigma}\cdot\vec{\ml{V}})\right]
\eea

We proceed to the third term of (\ref{fw}) which involves a double commutator of $\ml{O}$ with $\ml{E}$. Note that the single commutator produces
\bea
[\ml{O},\ml{E}]&=&[\ml{O}_\alpha,\ml{E}]+[\ml{O}_5,\ml{E}]\nn\\
&=&[\ml{O}_\alpha,\ml{E}_{VA}]+[\ml{O}_\alpha,\ml{O}_{\alpha5}]+[\ml{O}_5,\ml{E}_{VA}]+[\ml{O}_5,\ml{O}_{\alpha5}]
\eea
All terms except the momentum operator piece of $\ml{O}_\alpha$ contain a factor of $G_F$. Since we retain terms only up to $O(G_F)$,
\bea
[\ml{O},\ml{E}]&\approx&[ \vec{\alpha}\cdot\vec{p},\ \ml{E}_{VA}]+[\vec{\alpha}\cdot\vec{p} ,\ \ml{O}_{\alpha5}]\\
&=& \left(\frac{g_R+g_L}{g_R-g_L}\right) \vec{\alpha}\cdot(\vec{p}\ \ml{A})-\left(\frac{g_R-g_L}{g_R+g_L}\right) \vec{\alpha}\cdot\vec{p}\ (\vec{\sigma}\cdot\vec{\ml{V}})\nn
\eea
However, we want $[\ml{O}, [\ml{O},\ml{E}]]$ and 
\bea
 [\ml{O}, [\ml{O},\ml{E}]]&\approx&[ \ \vec{\alpha}\cdot\vec{p},\ [\ml{O},\ml{E}]]\\
&=& -\left[\left(\frac{g_R+g_L}{g_R-g_L}\right)\nabla^2 \ml{A}-\left(\frac{g_R-g_L}{g_R+g_L}\right) \nabla^2(\vec{\sigma}\cdot\vec{\ml{V}})\right]\nn
\eea
where we used $\alpha_i\alpha_j=(\delta_{ij}+i\epsilon_{ijk}\sigma_k)\mathbbm{1}$, which produces $(\vec{\alpha}\cdot\vec{p})^2=\alpha_i\alpha_j p_ip_j=p_ip_i=p^2=-\nabla^2$. Finally, consider the last term in Eq.(\ref{fw}),
\bea
 [\ml{O},\dot{\ml{O}}]&=& [\ml{O}_\alpha,\dot{\ml{O}_\alpha}]+ [\ml{O}_\alpha,\dot{\ml{O}_5}]+ [\ml{O}_5,\dot{\ml{O}_\alpha}]+ [\ml{O}_5,\dot{\ml{O}_5}]\nn\\
 &\approx&[\ml{O}_\alpha,\dot{\ml{O}_\alpha}]+ [\ml{O}_\alpha,\dot{\ml{O}_5}]+ [\ml{O}_5,\dot{\ml{O}_\alpha}]
\eea
Since $[\ml{O}_5,\dot{\ml{O}_5}]\sim O(G_F^2)$, we neglect this term. We evaluate the requisite commutators below, 
\bea
[\ml{O}_\alpha,\dot{\ml{O}_\alpha}] &=& [\vec{\alpha}\cdot(\vec{p}-\vec{\ml{V}}),-  \vec{\alpha}\cdot(\partial_t{\vec{\ml{V}}})]\nn\\
&\approx& -\vec{p}\cdot\partial_t\vec{\mathcal{V}} +i\vec{\sigma}\cdot\left(2\partial_t\vec{\mathcal{V}}\times\vec{p}-\vec{p}\times\partial_t\vec{\mathcal{V}}\right)
\\
\left[\ml{O}_\alpha,\dot{\ml{O}_5}\right]&=& [\vec{\alpha}\cdot(\vec{p}-\vec{\ml{V}}),\ \gamma^5\ml{\dot{A}}]\nn\\
 &\approx& \left[\vec{\alpha}\cdot\vec{p},\ \gamma^5\ml{\dot{A}}\right]=\vec{\sigma}\cdot\left(\vec{p} \,\ml{\dot{A}}\right)
\\
\left[\ml{O}_5,\dot{\ml{O}_\alpha}\right]&=& [\gamma^5\ml{A},\ \vec{\alpha}\cdot(-\partial_t\vec{\ml{V}})] = \ml{O}\left(G_F^2\right)
\eea
Thus, to the leading order in $G_F$ (which is $\ml{O}(G_F)$) we find
\bea
 [\ml{O},\dot{\ml{O}}]\approx -\vec{p}\cdot\partial_t\vec{\mathcal{V}} +i\vec{\sigma}\cdot\left(2\partial_t\vec{\mathcal{V}}\times\vec{p}-\vec{p}\times\partial_t\vec{\mathcal{V}}-i\vec{p}\,\dot{\ml{A}}\right)
\eea
We can now write down the interaction hamiltonian operator,
\bea
\hat{\ml{H}}^{(3)}_{e^{-}} &\approx& \beta\left[m_e-\frac{\left[\nabla^2+2i(\vec{\ml{V}}\cdot\vec{\nabla})+i(\vec{\nabla}\cdot\vec{\ml{V}})+\vec{\sigma}\cdot\left(\vec{\nabla}\times\vec{\ml{V}}\right)\right]+i \vec{\sigma}\cdot\left[2\ml{A}\ \vec{\nabla}+ (\vec{\nabla}\ml{A})\right]}{2m_e}\right]\nn\\
&&-\beta\ \frac{p^4}{8m_e^3}+\left[\left(\frac{g_R+g_L}{g_R-g_L}\right)\ml{A}-\left(\frac{g_R-g_L}{g_R+g_L}\right)(\vec{\sigma}\cdot\vec{\ml{V}})\right]\\
&&-\frac{1}{8m_e^2}\left[\left(\frac{g_R+g_L}{g_R-g_L}\right)\nabla^2 \ml{A}-\left(\frac{g_R-g_L}{g_R+g_L}\right) \nabla^2(\vec{\sigma}\cdot\vec{\ml{V}})\right]\nn\\
&&-\frac{i}{8m_e^2}\left[-\vec{p}\cdot\partial_t\vec{\mathcal{V}} +i\vec{\sigma}\cdot\left(2\partial_t\vec{\mathcal{V}}\times\vec{p}-\vec{p}\times\partial_t\vec{\mathcal{V}}-i\vec{p}\,\dot{\ml{A}}\right)\right]\nn
\eea

The fermion fields are quantized in a fiducial box of volume $V$ which cancels out in calculating any physical quantity. The field operator is expanded in the standard fashion using creation and annhilation operators as
\bea\label{field}
\hat{\psi}(x,t)=\sum_k\sum_{s=1,2}\sqrt{\frac{m_e}{EV}}\left[b^{(s)}_ku^{(s)}(k)e^{ik\cdot x}+d^{(s)\dagger}_kv^{(s)}(k)e^{-ik\cdot x}\right]
\eea
where $b^{(s)}(k)$ and $d^{(s)}(k)$ satisfy the anti-commutation relations,
\bea
\left\{b_k^{(s)},b_{k'}^{(s')\dagger}\right\}=\delta_{ss'}\delta_{kk'},\ \left\{d_k^{(s)},d_{k'}^{(s')\dagger}\right\}=\delta_{ss'}\delta_{kk'}
\eea
and all other anti-commutators are zero. 

\subsection{Transition Probability, Cross-section \& Rate}
\label{app:transitionprobability}
The transition probability for the process $e^{-}+\nu_{e^{-}}\ra e^{-}+\nu_{e^{-}}$, with the electron field $e^{-}$ being those in the solid, impinged upon by an incident neutrino, can be calculated using Fermi's golden rule,
\bea
\Gamma = {2\pi}\ \Big|\bra{\vec{p}_2;i}\hat{H}_{e^-}\ket{\vec{p}_1;f}\Big|^2\delta(E_i-E_f+E_1-E_2)
\eea
where $\vec{p}_1$ and $\vec{p}_2$ are the incident and scattered momenta of the neutrinos, $i$ and $f$ are the initial and final solid electronic states, and similarly for $E_{1,2}$ and $E_{i,f}$, respectively. Note that in this appendix, we choose $(E_1,\ \vec{p}_1)$ and $(E_2,\  \vec{p}_2)$ instead of $(E_{\nu, i},\ \vec{p}_{\nu, i})$ and $(E_{\nu, f},\ \vec{p}_{\nu,f})$ to keep the notation general. The leading order term is
\bea
\hat{H}_{e^-}&\supset& \left[\left(\frac{g_R+g_L}{g_R-g_L}\right)\ml{A}-\left(\frac{g_R-g_L}{g_R+g_L}\right)(\vec{\sigma}\cdot\vec{\ml{V}})\right]\nn\\
&=& \frac{G_F}{\sqrt{2}}\left[(g_R+g_L)\ (V^0-A^0)-(g_R-g_L)\ \vec{\sigma}\cdot(\vec{V}-\vec{A})\right]\nn\\
&\equiv& \frac{G_F}{\sqrt{2}}\left[g_+(V^0-A^0)-g_-\ \vec{\sigma}\cdot(\vec{V}-\vec{A})\right]
\eea
Denoting $g_+\equiv(g_R+g_L)$ and $g_-\equiv(g_R-g_L)$, we get three terms between in and out states,
\bea
g_{+}^2\ra\left(\frac{G_F}{\sqrt{2}}\right)^2\frac{g_{+}^2}{E_1E_2}\left[E_1E_2+\vec{p}_1\cdot\vec{p}_2\right] \frac{S_1(\omega, \vec{k})}{V^2}
\eea
\bea
g_{-}^2\ra\left(\frac{G_F}{\sqrt{2}}\right)^2\frac{g_{-}^2}{E_1E_2V^2}\left[iE_1\ (\vec{p}_2\cdot\vec{S}^\sigma_4)-iE_2\ (\vec{p}_1\cdot\vec{S}^\sigma_4)+2\ p_1^ip_2^j\ S^{ij,\sigma}_5\right.\nn\\
\left.+\ (E_1E_2-\vec{p}_1\cdot\vec{p}_2)\ S^\sigma_3\right]\hspace{32mm}
\eea
\bea
g_{+}g_{-}\ra\left(\frac{G_F}{\sqrt{2}}\right)^2\frac{g_{+}g_{-}}{E_1E_2V^2}\left[E_2\ (\vec{p}_1\cdot\vec{S}^\sigma_2)+E_1\ (\vec{p}_2\cdot\vec{S}^\sigma_2)+i\ \vec{S}^\sigma_2\cdot(\vec{p}_1\times\vec{p}_2)\right]
\eea
where $\omega\equiv E_1 - E_2$ is the energy deposited and $\vec{k} \equiv \vec{p}_1-\vec{p}_2$ is momentum transferred to the solid. Adding the complex conjugate,
\bea
g_{+}g_{-}+g_{-}g_{+}\ra\left(\frac{G_F}{\sqrt{2}}\right)^2\frac{g_{+}g_{-}}{E_1E_2V^2}\left[E_2\ (\vec{p}_1\cdot\vec{S}^\sigma_2)+E_1\ (\vec{p}_2\cdot\vec{S}^\sigma_2)\right]
\eea
We sum over all the electronic positions in the solid to calculate the double differential cross-section in addition to summing over the initial and final electronic states. Now, the dynamic structure factors, $S_i(\omega,  \vec{k})$ which are dependent on the properties of the crystal are defined as,
\bea
{S_1(\omega, \vec{k})} &\equiv& \sum_{if}\ \big|\sum_j\bra{f}e^{i(\vec{p}_2-\vec{p}_1)\cdot\vec{x}^j_e}\ket{i}\big|^2 \delta(E_i-E_f+\omega)\nn\\
&=&\sum_{i}\ \sum_{jj'}\bra{i}e^{-i\vec{k}\cdot(\vec{x}^j_e-\vec{x}^{j'}_e)}\ket{i} \delta(E_i-E_f+\omega)
\\
{\vec{S}^\sigma_2(\omega, \vec{k})} &\equiv& \sum_{if}\ \sum_{jj'}\bra{i}\vec{\sigma}_e^{j'}\ e^{-i(\vec{p}_2-\vec{p}_1)\cdot\vec{x}^{j'}_e}\ket{f}\bra{f}e^{i(\vec{p}_2-\vec{p}_1)\cdot\vec{x}^j_e}\ket{i} \delta(E_i-E_f+\omega)\nn\\
&=&\sum_{i}\ \sum_{jj'}\bra{i}e^{-i\vec{k}\cdot(\vec{x}^j_e-\vec{x}^{j'}_e)}\vec{\sigma}_e^{j'}\ket{i} \delta(E_i-E_f+\omega)
\\
{S^\sigma_3(\omega, \vec{k})} &\equiv& \sum_{if}\ \sum_{jj'}\bra{i}\vec{\sigma}_e^{j'}e^{-i(\vec{p}_2-\vec{p}_1)\cdot\vec{x}^{j'}_e}\ket{f}\cdot\bra{f}\vec{\sigma}_e^je^{i(\vec{p}_2-\vec{p}_1)\cdot\vec{x}^j_e}\ket{i}\ \delta(E_i-E_f+\omega)\nn\\
&=&\sum_{i}\ \sum_{jj'}\bra{i}e^{-i\vec{k}\cdot(\vec{x}^j_e-\vec{x}^{j'}_e)}\vec{\sigma}_e^{j'}\cdot\vec{\sigma}_e^{j}\ket{i} \delta(E_i-E_f+\omega)
\\
{\vec{S}^\sigma_4(\omega, \vec{k})} &\equiv& i\sum_{if}\ \sum_{jj'}\bra{i}\vec{\sigma}_e^{j'}e^{-i(\vec{p}_2-\vec{p}_1)\cdot\vec{x}^{j'}_e}\ket{f}\times\bra{f}\vec{\sigma}_e^je^{i(\vec{p}_2-\vec{p}_1)\cdot\vec{x}^j_e}\ket{i}\delta(E_i-E_f+\omega)\nn\\
&=&i\sum_{i}\ \sum_{jj'}\bra{i}e^{-i\vec{k}\cdot(\vec{x}^j_e-\vec{x}^{j'}_e)}\vec{\sigma}_e^{j'}\times\vec{\sigma}_e^{j}\ket{i} \delta(E_i-E_f+\omega)
\\
{S^{mn, \sigma}_5(\omega, \vec{k})} &\equiv& \sum_{if}\ \sum_{jj'}\bra{i}\sigma_e^{m,j'}\ e^{-i(\vec{p}_2-\vec{p}_1)\cdot\vec{x}^{j'}_e}\ket{f}\bra{f}\sigma_e^{n,j}e^{i(\vec{p}_2-\vec{p}_1)\cdot\vec{x}^j_e}\ket{i}\delta(E_i-E_f+\omega)\nn\\
&=&\sum_{i}\ \sum_{jj'}\bra{i}e^{-i\vec{k}\cdot(\vec{x}^j_e-\vec{x}^{j'}_e)}\sigma_e^{m,j'}\sigma_e^{n,j}\ket{i} \delta(E_i-E_f+\omega)
\eea
where $\vec{\sigma}_e$ are the $SU(2)$ Pauli matrices. Note that besides $S_1(\omega, \vec{k})$, all the other structure factors are dependent on the electron spin matrices. Now that kinematics of the interaction dictates that, for a fixed incident energy $E_1$, the variables in a $2\ra2$ inelastic scattering process are $\vec{k}$, $\omega$, and angle of scattering in the plane, $\theta$. However, these are not all independent. By energy-momentum conservation:
\bea\label{q_kin}
k^2 &=& p_1^2+p_2^2-2\vec{p}_1\cdot\vec{p}_2
\implies\cos\theta = \frac{E_1^2+(E_1-\omega)^2-k^2}{2E_1(E_1-\omega)}
\eea
The delta functions embedded in each of the dynamic structure factors above impose energy conservation, implying that only $E_f-E_i=\omega$ is allowed.\\

To calculate the cross-section from the transition probability, we multiply by the density of final (scattered) neutrino states, $\rho(E_2)$, and divide by the incoming neutrino flux $I_\nu$,
\bea
\frac{d^2\sigma}{dE_2d\Omega}&=&\frac{\Gamma\rho(E_2)}{I_\nu}
\eea
Using the relativistic energy-momentum relation, $E=\sqrt{\omega_k^2+m_e^2}$, and that the density of states in a fiducial quantization volume $V$ is given by
\bea
\rho(k)dk=\frac{V}{(2\pi)^3}k^2dk\implies \frac{\rho(E_2)}{I_\nu}=\frac{V^2}{(2\pi)^3}{E_2^2}
\eea
we arrive at the double differential cross-section,
\bea
\frac{d^2\sigma}{dE_2d\Omega}=\frac{V^2E_2^2}{(2\pi)^2}\ \Gamma
\eea
where
\bea
\label{eq:M_NR2}
\left(\frac{E_1E_2V^2}{4G_F^2}\right)\Gamma&=&g_{+}^2\left(E_1E_2+\vec{p}_1\cdot\vec{p}_2\right) S_1+ 2 g_{+}g_{-}\left\{E_2\ (\vec{p}_1\cdot\vec{S}^\sigma_2)+E_1\ (\vec{p}_2\cdot\vec{S}^\sigma_2)\right\}\\
&+&g_{-}^2\left(E_1\ (\vec{p}_2\cdot\vec{S}^\sigma_4)-E_2\ (\vec{p}_1\cdot\vec{S}^\sigma_4)+2\ p_1^ip_2^j\ S^{ij,\sigma}_5+ (E_1E_2-\vec{p}_1\cdot\vec{p}_2)\ S^\sigma_3\right)\nn
\eea
The three independent kinematic variables in the plasmon production (or any inelastic) process are - $\{E_1, \omega, k\}$. This follows from energy and momentum conservation due to: (i) $E_2=E_1-\omega\implies\omega =E_f-E_i$, (ii) $E_2$ in turn gives us $p_2$ whereas, (iii) $E_1$ directly gives us $p_1$, and (iv) using Eq.(\ref{q_kin}) and $k$, we can get $\cos\theta$ which is the (planar) scattering angle, thus completely determining the kinematic variables. Therefore, we find for a fixed incident energy $E_1$, deposited energy $\omega$, and momentum transfer $k$, that the cross-section can be written as,
\bea
\frac{d^2\sigma}{dE_2d\Omega}\Bigg|_{\rm fixed\ \{k,\ E_1,\ \omega\}}=\frac{V^2(E_1-\omega)^2}{(2\pi)^2}\Gamma
\label{eq:d2sigmadE2dOmega}
\eea

The total scattering rate for a particle of number density $n_1$, velocity $v_1$, and scattering cross-section $\sigma$  is given by
\bea
\ml{R} = (n_1 v_1)\sigma = \sigma\Phi_1
\eea
where the combination $n_1v_1=\Phi_1$ is the incident particle flux. If this flux is energy dependent, then the product becomes an integral over the incident energy,
\bea
\Phi_1 = n_1v_1 = \int dE_1 \frac{d\Phi_1}{dE_1}
\eea
Moreover, in the present case of neutrino scattering resulting in plasmon production, we see from (\ref{eq:M_NR2}) that the scattering cross-section is energy dependent, $\sigma\equiv\sigma(E_1)$. Thus,
\bea\label{R}
\ml{R}  &=& \int \frac{d\Phi_1}{dE_1}\sigma(E_1)dE_1\nn\\
&=&\int \frac{d\Phi_1}{dE_1}\left[\int\frac{d^2\sigma}{ dE_2 d\Omega}\ dE_2 d\Omega\right]_{\rm fixed\ E_1}dE_1
\eea
We can calculate the differential rate evaluated at a specific recoil energy $\omega_0$ using the following $\delta$-function insertion,
\bea
\frac{d\ml{R}}{d\omega}\Bigg|_{\omega_0}=\int \frac{d\Phi_1}{dE_1}\left[\int\frac{d^2\sigma}{ dE_2 d\Omega}\ dE_2 d\Omega\right]_{\rm fixed\ E_1}dE_1\ \delta(\omega-\omega_0)
\eea
where $\delta(\omega-\omega_0)$ is inside the integral. This can be verified by integrating both sides with respect to $\omega_0$ which gives back Eq.(\ref{R}) when $\omega$ corresponds to the recoil energy determined by the set $\{E_1, E_2, k\}$. We can absorb this $\delta$-function inside the $[...]$ integral, simplifying the $E_2$ integration by the substitution $E_2\ra E_1-\omega$ (enforced by the $\delta$-functions embedded in the definition of the dynamic structure factors) in the expression for the double-differential cross-section,
\bea
\frac{d\ml{R}}{d\omega}\Bigg|_{\omega_0}=\int \frac{d\Phi_1}{dE_1}\left[\int\frac{d^2\sigma}{ dE_2 d\Omega}\Bigg|_{\rm fixed\ \{k, E_1, E_2=E_1-\omega_0\}} d\Omega\right]_{\rm fixed\ \{E_1, \omega_0\}}dE_1\
\eea

Using this, we can calculate the differential rate with respect to the recoil energy,
\bea\label{diff_rate}
\frac{d\ml{R} }{d\omega} =\int_{E_{1,\rm{min}}} \frac{d\Phi}{dE_1}\left[\int d\Omega\left\{\frac{d^2\sigma}{dE_2d\Omega}\right\}_{\rm fixed\ \{k, E_1, \omega\}}\right]_{\rm fixed\ \{E_1,\ \omega\}} dE_{1}\Bigg|_{\rm fixed\ \omega}
\eea
where $d\Phi/dE_1$ is the differential flux of the incident neutrinos (since the flux is energy-dependent). The measure for the integral in $[...]$ can be further simplified using Eq.(\ref{q_kin})
\bea
d\Omega=2\pi d(\cos\theta)=-\frac{k}{E_1(E_1-\omega)}dk
\eea
Using Eq.(\ref{eq:d2sigmadE2dOmega}) the differential rate at a fixed $\omega=\omega_0$ thus becomes
\bea
\frac{d\ml{R} }{d\omega}\Bigg|_{\omega_0}=-\int_{E_{1,\rm{min}}} \frac{d\Phi}{dE_1}\left[\int_{k_{min}}^{k_{max}} dk\ \frac{k\Gamma'}{2\pi}\right](E_1-\omega)\frac{dE_1}{E_1}\Bigg|_{\rm fixed\ \omega=\omega_0}
\eea
where $\Gamma'=V^2\Gamma$ and $\Gamma$ is given by Eq.(\ref{eq:M_NR2}). \\

\section{Plasmon Bremsstrahlung by Neutrinos} 
\label{app:brem}

We consider the generation of plasmons in the solid via a nuclear bremsstrahlung process whereby the neutrino impinges upon the nucleus and the plasmon mode results from the electron being given an impulse during this process. The resulting density fluctuations in the solid manifest themselves as a plasmon mode. To calculate the rate of plasmon production via bremsstrahlung, we need to posit an interaction hamiltonian between the electrons and the fixed ions in the solid. We assume a homogeneous electron gas with linear screening of the interaction between nuclei and electrons. Under this assumption, it can be shown~\cite{mah00} that the interaction can be written in a self-consistent manner in terms of a frequency-dependent constant, $\epsilon(\omega,\vec{r},\vec{r'})$ (called the dielectric constant) which modifies the usual Coulomb interaction between the charges. The dielectric constant thus `renormalizes' or absorbs the leading order effects of screening in an electron gas background. This allows us to write the interaction Hamiltonian between the electron gas and a positively charged ion,
\bea
\ml{H}_{int}^{eN}(\vec{r})=-\int d\vec{r'}\frac{Z\alpha}{\epsilon(\omega,\vec{r},\vec{r'})}\frac{1}{|\vec{r'}-\vec{r}_N|}
\eea
where $Z$ is the number of protons in the positively charged ion cores of the solid crystal, $\alpha$ is the fine-structure constant (working in natural units) and $\vec{r}_N$ is the location of a positively charged ion. Here $\vec{r}$ is the location of the electron wave-function where the interaction hamiltonian density operator is desired. The dielectric constant can be calculated solely in terms of the density fluctuations of the free electrons in the solid and as such is a meaningful quantity without an application of an external electric field (though an adiabatic process of turning on the external field is required to define it). In general the dielectric function is a tensor object but we assume that (i) the dielectric matrix is diagonal, (ii) under the assumption of distances being much larger than the lattice sizing, the dielectric constant carries the translational symmetry of the lattice and (iii) the solid is homogeneous so that all the diagonal elements in (i) are the same.

Since the asymptotic states are of fixed momentum, we re-write the interaction hamiltonian between electrons and the fixed ions in terms of the Fourier transform of the dielectric constant, $\epsilon(\omega, \vec{k}$)
\bea\label{solid_hamiltonian}
\ml{H}_{int}^{eN}=-\frac{4\pi Z\alpha}{V}\sum_{k}\frac{e^{i\vec{k}\cdot(\vec{r}-\vec{r}_N)}}{k^2\epsilon(\omega, \vec{k})}
\eea
Note that we imagine the system to be in a box of volume $V$ (for normalization purposes) and so we deal with only discrete values of $k$. The total hamiltonian also includes the interaction between the neutrino and the nucleus, $\ml{H}^{\nu N}_{int}$  which is the same as the electron-neutrino Hamiltonian in (\ref{e_nu_op}) except that now $g_R$ and $g_L$ are now different depending on the quarks and we need to sum over both the $u$ and $d$ quarks. The coefficients for each are given 
\bea
g_L^u&=&\rho^{NC}_{\nu N}\left(\frac{1}{2}-\frac{2}{3}\hat{\kappa}_{\nu N}\sin^2\theta_w\right)+\lambda_L^u\nn\\
g_L^d&=&\rho^{NC}_{\nu N}\left(-\frac{1}{2}+\frac{1}{3}\hat{\kappa}_{\nu N}\sin^2\theta_w\right)+\lambda_L^d\nn\\
g_R^u&=&\rho^{NC}_{\nu N}\left(-\frac{2}{3}\hat{\kappa}_{\nu N}\sin^2\theta_w\right)+\lambda_R^u\nn\\
g_R^d&=&\rho^{NC}_{\nu N}\left(\frac{1}{3}\hat{\kappa}_{\nu N}\sin^2\theta_w\right)+\lambda_R^d
\eea
where $\rho^{NC}_{\nu N}=1.0086$, $\hat{\kappa}_{\nu N}=0.9978$, $\lambda_L^{u}=-0.0031$, $\lambda_L^{d}=-0.0025$ and $\lambda_R^d=2\lambda_R^u=7.5\times10^{-5}$. The total Hamiltonian then is
\bea
\ml{H}=\ml{H}_{int}^{eN}+\ml{H}^{\nu N}_{int}
\eea

\begin{figure*}[t]
    \centering
    \includegraphics[width=7cm]{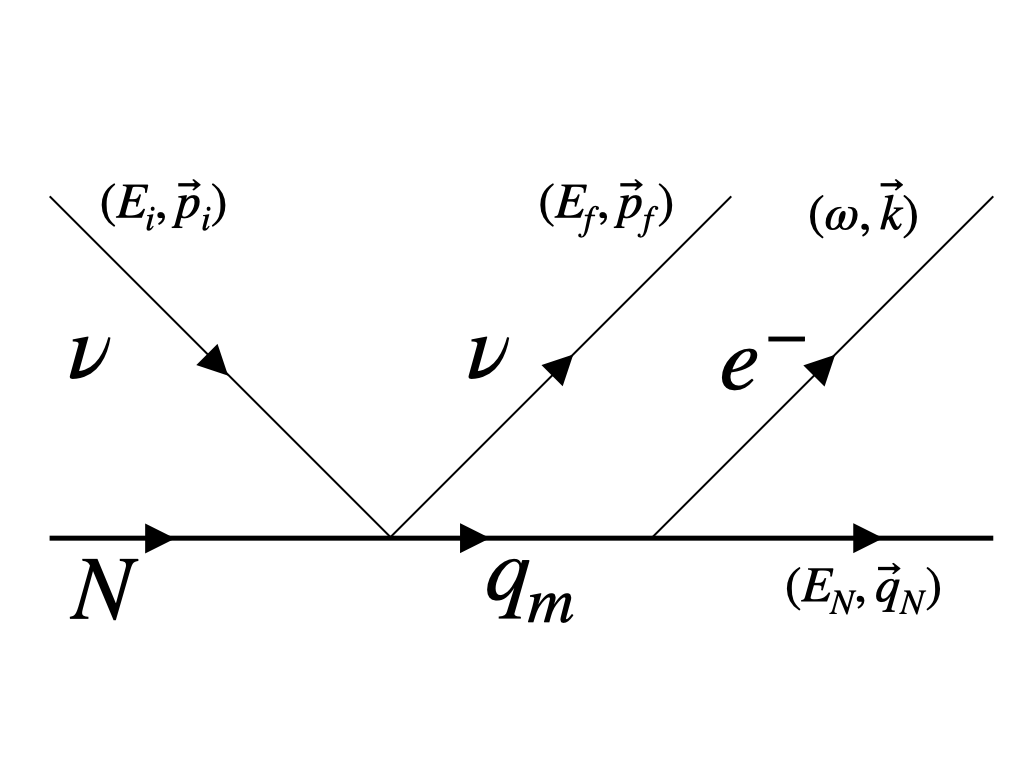}\hspace{5mm} 
    \includegraphics[width=7cm]{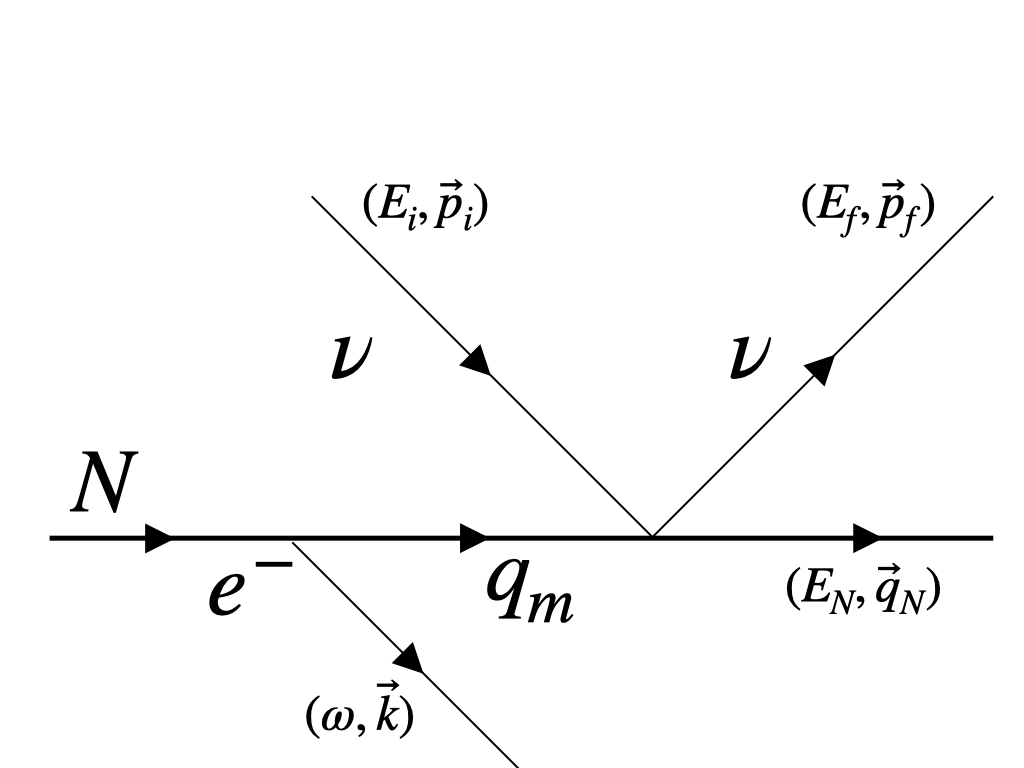}\\
    \caption{The $s$-channel (left) and $u$-channel (right) diagrams that contribute to the bremsstrahlung process.}
    \label{fig:FeynmanDiagrams}
\end{figure*}

We modify previously used notation by defining $(\omega, \vec{k})_{1,2}$ as the initial and final energy/momenta for the electrons in the solid and $(E, \vec{p})_{1,2}$ as the initial and final energy/momenta for the neutrinos respectively. The positive ion will be considered at rest initially ($\epsilon_N=m_N$) and its final energy momentum is denoted as $(E_N, \vec{q}_N)$. In Fig.~\ref{fig:FeynmanDiagrams} are the two diagrams, following the setup of~\cite{Knapen:2020aky}, that contribute to the bremsstrahlung process -- (i) an $s$-channel diagram where the electron is scattered after the scattering between neutrinos and the nucleus is `complete' and (ii) a $u$-channel diagram where the electron is scattered first. This process is described by second order perturbation theory (refer to the figure) and the total amplitude for the process is given as
\bea
\ml{M}=\ml{M}_s+\ml{M}_u
\eea 
where $\ml{M}_{1,2}$ correspond to the $s$ and $u$ channel diagrams respectively. In second order perturbation theory, the transition amplitude depends on all possible intermediate states,
\bea
\ml{M}=\sum_m\frac{\bra{out}\ml{H}\ket{m}\bra{m}\ml{H}\ket{in}}{\epsilon_{in}-\epsilon_{m}}
\eea
where $\ket{m}$ is the set of all intermediate states for the process and $\{\ket{in}, \ket{out}\}$ are the in and out states consistent with the intermediate state $\ket{m}$ for a plasmon bremsstrahlung process. Note that energy is not conserved at the vertex level in second order perturbation theory. But overall energy is still conserved in the process,
\bea\label{energy}
\epsilon_N+E_1+\omega_1=E_N+E_2+\omega_2
\eea

Now, the $s$-channel diagram has the neutrino and at-rest solid nucleus as the in-state, the scattered neutrino and the scattered nucleus in the intermediate state, and the out state includes the nucleus and a scattered electron. Therefore, the denominator for the $s$-channel diagram is, 
\bea
\epsilon_{in}-\epsilon_{m}=(\epsilon_N+E_1)-(E_2+E_q)
\eea

Since the ions are non-relativistic,
\bea
\epsilon_N&=&m_N\nn\\
E_N&=&m_N+\frac{q_N^2}{2m_N}\\
E_q&=&m_N+\frac{q^2}{2m_N}\nn
\eea

Even though energy isn't conserved at the vertex, the nature of interaction and fixed-momentum in and out states imply that momentum is still conserved at each vertex (this follows from the the interaction hamiltonian and the in-out states in the numerator of ),
\bea\label{s_momentum}
\Delta\vec{k} \equiv\vec{k}_2-\vec{k}_1=\vec{q}-\vec{q}_N
\eea
Taken together (\ref{energy})-(\ref{s_momentum}) imply,
\bea
\epsilon_{in}-\epsilon_{m}=\Delta\omega(\equiv\omega_2-\omega_1)-\frac{\Delta\vec{k}\cdot\vec{q}_N}{m_N}-\frac{\vec{q}_N^2}{2m_N}\ra\mbox{$s$-channel}
\eea

On the other hand in the $u$-channel diagram, the neutrino is replaced by the electron in the in-state, whereas the out state now include the nucleus and a scattered neutrino. Thus, 
\bea
\epsilon_{in}-\epsilon_{m}=(\epsilon_N+\omega_1)-(\omega_2+E_q)
\eea
and momentum conservation at the first vertex gives,
\bea
\vec{q}=-\Delta\vec{k}
\eea
together implying
\bea
\epsilon_{in}-\epsilon_{m}=-\left(\Delta\omega+\frac{\Delta \vec{k}^2}{2m_N}\right)\ra\mbox{$u$-channel}
\eea

With the aid of the specified in and out states and the denominators, we can write down the matrix elements of the two diagrams (note that $\ml{H}_{int}^{eN}$ acts on the Hilbert space containing electron+nucleus whereas $\ml{H}^{\nu N}_{int}$ acts on the Hilbert space containing neutrino+nucleus)
\bea
\ml{M}_s=\sum_{\vec{q}}\frac{\bra{\vec{q}_N;\vec{k}_2}\ml{H}_{int}^{eN}\ket{\vec{q};\vec{k}_1}\bra{\vec{q};\vec{p}_2}\ml{H}^{\nu N}_{int}\ket{0;\vec{p}_1}}{\Delta\omega-{\Delta\vec{k}\cdot\vec{q}_N}/{m_N}-{\Delta\vec{k}^2}/{2m_N}}
\eea
\bea
\ml{M}_u=-\sum_{\vec{q}}\frac{\bra{\vec{q}_N;\vec{p}_2}\ml{H}_{int}^{\nu N}\ket{\vec{q};\vec{p}_1}\bra{\vec{q};\vec{k}_2}\ml{H}^{e N}_{int}\ket{0;\vec{k}_1}}{\Delta\omega+{\Delta \vec{k}^2}/{2m_N}}
\eea
Summing over all the intermediate states $\{\ket{m}\}$ is equivalent to summing over all momenta $\vec{q}$ of the intermediate state. In the case of $\ml{M}_s$, the numerator has a term like $\bra{\vec{q}_N;\vec{k}_2}\ml{H}_{int}^{eN}\ket{\vec{q};\vec{k}_1}$ which for the electronic matrix element is $\bra{\vec{q}_N}e^{-i\vec{k}\cdot\vec{r}_N}\ket{\vec{q}}=\delta^3(\vec{q}-\vec{k}-\vec{q}_N)$ and so the sum over $\vec{q}$ just enforces $\vec{q}=\vec{k}+\vec{q}_N$. Furthermore, in the case of electrons in Bloch states, $\bra{\vec{k}_2}e^{i\vec{k}\cdot\vec{r}}\ket{\vec{k}_1}$ is non-zero only if $\vec{k}\equiv\vec{k}_2-\vec{k}_1$. Thus, both sums over $\vec{q}$ and $\vec{k}$ are eliminated in exchange for $\vec{q}=\Delta\vec{k}+\vec{q}_N$ in $\bra{\vec{q}_N;\vec{p}_2}\ml{H}_{int}^{\nu N}\ket{\vec{q};\vec{p}_1}$. And the neutrino/nuclear matrix element, $\bra{\vec{q};\vec{p}_2}\ml{H}^{\nu N}_{int}\ket{0;\vec{p}_1}$ simplifies to $\bra{\vec{k}+\vec{q}_N;\vec{p}_2}\ml{H}^{\nu N}_{int}\ket{0;\vec{p}_1}$\\

In the case of $\ml{M}_u$ on the other hand the electronic matrix element enforces $\delta^3(\vec{q}+\vec{k})$. Thus, the neutrino/nuclear matrix element can be written as $\bra{\vec{q}_N;\vec{p}_2}\ml{H}_{int}^{\nu N}\ket{-\vec{k};\vec{p}_1}$, which is shifted with respect to $\bra{\vec{k}+\vec{q}_N;\vec{p}_2}\ml{H}^{\nu N}_{int}\ket{0;\vec{p}_1}$ by the fact that the in and out states of the nucleus differ by an inertial frame. Such a boost keeps the matrix element for a local momentum conserving interaction invariant.\\

The relativistic interaction between the nucleus and neutrinos given by $\ml{H}_{int}^{\nu N}$ conserves momentum, providing overall momentum conservation as expected. The total matrix element can now be written as (for fixed in and out momenta)
\bea
\ml{M}=\sum_{\vec{k}}\frac{4\pi \alpha Z \ml{M}_{\nu N}\bra{\vec{k}_2}e^{i\vec{k}\cdot\vec{r}}\ket{\vec{k}_1}}{Vk^2\epsilon(\Delta\omega, \vec{k})}\left[\frac{1}{\Delta\omega-{\vec{k}\cdot\vec{q}_N}/{m_N}-{\vec{k}^2}/{2m_N}}-\frac{1}{\Delta\omega+{ \vec{k}^2}/{2m_N}}\right]\nn\\
\eea

To translate the matrix element into a cross-section, we need to sum over all the initial states and average over final states. Also, we let $\vec{k}$ denote $\Delta\vec{k}$ in accordance with momentum conservation imposed by Bloch states. If we consider the soft plasmon limit (analogous to soft  photon emission), $\vec{k}\cdot\vec{q}_N\ll\Delta\omega$ and $k\ll q_N$, where $\vec{k}\equiv\Delta\vec{k}$ for Bloch electron states, then
\bea
\ml{M}=\sum_{\vec{k}}\frac{4\pi \alpha Z \ml{M}_{\nu N}\bra{\vec{k}_2}e^{i\vec{k}\cdot\vec{r}}\ket{\vec{k}_1}|\vec{v}_{ion}|\cos\theta}{Vk\Delta\omega^2 \epsilon(\Delta\omega, \vec{k})}
\eea
where $\vec{v}_{ion}=\vec{q}_N/m_N$ is the recoil velocity of the ion and $\theta$ is the angle between $\vec{q}_N$ and $\hat{k}$. We would like to calculate the double differential cross-section, $d^2\sigma/dE_Rdk$ where $E_R=q_N^2/(2m_N)$ is the recoil energy of the ion. To do this, we first need the rate for the bremsstrahlung process, for which we need to average the squared matrix element over all initial states and sum over all final states of the scattered neutrino. Thus, we sum over $\vec{k}_1$ and $\vec{k}_2$. To connect the electronic matrix element with the structure factors and dielectric constant, we introduce an (redundant) integral over Dirac delta function, $\delta(\omega-\Delta\omega)$
\bea
&&\frac{1}{4}\int d\omega\sum_{\vec{k}_1,\vec{k}_2, \vec{k}}|\ml{M}|^2\delta(\omega-\Delta\omega)\\
&&=\int d\omega\sum_{\vec{k}} \frac{4\alpha Z^2|\vec{v}_{ion}|^2\cos^2\theta}{4\Delta\omega^2|\epsilon(\Delta\omega, \vec{k})|^2}\frac{|\ml{M}_{\nu N}|^2}{V}\left[\frac{4\pi^2\alpha}{k^2V}\sum_{\vec{k}_1}\big|\bra{\vec{k}_1+\vec{k}}e^{i\vec{k}\cdot\vec{r}}\ket{\vec{k}_1}\big|^2\delta(\omega-\Delta\omega)\right]\nn
\eea
Using the nature of momentum conserving matrix elements for Bloch electrons in the solid, we can take $\vec{k}_1$ and $\vec{k}$ to be the independent momentum variables where the Dirac delta function $\delta^3(\vec{k}_1-\vec{k}-\vec{k}_2)$ of Bloch matrix elements is eliminated in the structure function. Notice then the expression in the brackets is $\Im[\epsilon(\omega, \vec{k})]$~\cite{Dressel2002Electrodynamics}. The absence of the Pauli blocking factor, $f(\omega_1,T)[1-f(\omega_2,T)]$ in our expression is due to the fact that we are working at zero temperature where for occupation number density $f$ defined by the Fermi-Dirac distribution, $\lim_{T\ra0}f_{FD}(\omega_1,T)[1-f_{FD}(\omega_2,T)]=1$.

Next to do the $d^3k$ integrals, we define the solid angle $\theta$ with respect to $\vec{q}_N$,
\bea
&&\frac{1}{4}\int d\omega\sum_{\vec{k}_1,\vec{k}_2, \vec{k}}|\ml{M}|^2\delta(\omega-\Delta\omega)\\
&&=\frac{1}{4}\int d\omega\int 4\frac{dk}{(2\pi)^3}\ \pi k^2 d(\cos\theta) \frac{4\alpha Z^2|\vec{v}_{ion}|^2\cos^2\theta}{\omega^4\pi}{|\ml{M}_{\nu N}|^2}\ \Im\left[-\frac{1}{\epsilon(\omega, \vec{k})}\right]\nn\\
&&= \frac{16\pi\alpha Z^2|\vec{v}_{ion}|^2}{3}\int \frac{dk}{(2\pi)^3}\int d\omega{|\ml{M}_{\nu N}|^2}\frac{k^2}{\omega^4}\ \Im\left[-\frac{1}{\epsilon(\omega, \vec{k})}\right]\nn
\eea
In going from $\sum_{\vec{k}}/V$ to $\int d^3k$, we need to divide by the volume of phase space $(2\pi)^3$. To get $d^2\sigma/dE_fd\Omega$, we multiply the above with $2\pi$ (for rate) and density of final neutrino states and divide by the incident flux. We also employed the complex identity $\Im[\epsilon(\omega, \vec{k})]/|\epsilon(\omega, \vec{k})|^2=\Im[-1/\epsilon(\omega, \vec{k})]$. We then have (using $|v_{ion}|=2E_R/m_N$)
\bea
\frac{d^2\sigma}{dE_fd\Omega}=\frac{16\pi\alpha Z^2}{3}\frac{2E_R}{m_N}\int \frac{dk}{(2\pi)^3}\int d\omega\left[2\pi\frac{|\ml{M}_{\nu N}|^2\rho(E_f)}{4I_\nu}\right]\frac{k^2}{\omega^4}\ \Im\left[-\frac{1}{\epsilon(\omega, \vec{k})}\right]\nn\\
\eea
where $E_f$ is the final state energy of the ions electrons/plasmons combined and $d\Omega$ is the differential solid angle of the scattered neutrino. Note that the term in $[...]$ on the right-hand-side is the double differential cross-section for elastic scattering between the neutrino and the nucleus, $d^2\sigma_{\nu N}/dE_fd\Omega$ . $V^2$, coming from $\rho(E_f)/I_\nu$, will be canceled by the quadratic nature of the field operators in the interaction each of which produces a factor of $1/\sqrt{V}$ as in the right-hand-side of (\ref{field}). Thus we have,
\bea
\frac{d^2\sigma}{dE_fd\Omega}\Bigg|_{E_R^0}=\frac{32\pi\alpha Z^2}{3}\frac{dk}{(2\pi)^3}\int d\omega\frac{E_R^0}{m_N}\left[\frac{d^2\sigma_{\nu N}}{dE_fd\Omega}\right]\frac{k^2}{\omega^4}\ \Im\left[-\frac{1}{\epsilon(\omega, \vec{k})}\right]
\eea
Integrating both sides with respect to $E_f$ and $\Omega$, after multiplying by $\delta(E_R-E_R^0)$, gives
\bea
\frac{d\sigma}{dE_R}\Bigg|_{E_R^0}=\frac{4Z^2\alpha}{3\pi^2}\frac{E_R^0}{m_N}\int dk\int d\omega\left[\frac{d\sigma_{\nu N}}{dE_R}\right]_{E_R=E_R^0+\omega}\frac{k^2}{\omega^4}\ \Im\left[-\frac{1}{\epsilon(\omega, \vec{k})}\right]
\eea
The neutrino-nucleus cross-section on the right-hand-side has the same in and out neutrino energies. Therefore, the nuclear recoil energy in evaluating the differential cross-section $d\sigma_{\nu N}/dE_R$ is $E_1-E_2=E_R+\Delta\omega$, where $\Delta\omega$ is the energy deposited in the electrons (because of the Dirac delta function $\delta(\omega-\Delta\omega)$, the recoil energy inside the $d\omega$ integral is $E_R+\omega$ where $E_R$ is the actual recoil energy of the nucleus in the brem process). Finally, we can further simplify to obtain the double differential cross-section over nuclear recoil energy $E_R$ and the energy deposited $\omega$,
\bea
\frac{d^2\sigma}{dE_Rd\omega}=\frac{4Z^2\alpha}{3\pi^2}\frac{E_R}{m_N}\int dk \frac{k^2}{\omega^4}\left[\frac{d\sigma_{\nu N}}{dE_R}\right]\Im\left[-\frac{1}{\epsilon(\omega, \vec{k})}\right].
\eea
In practice the potential felt by the valence electrons is screened by the tightly bound core electrons. Following~\cite{Knapen:2020aky}, we replace $Z$ with $Z_{\rm{ion}}(k)$, the Fourier transformed potential which includes the charge screening effects.

\bibliography{plasmon}

\end{document}